\documentclass[fleqn,usenatbib]{mnras}

\usepackage{newtxtext,newtxmath}
\usepackage[T1]{fontenc}

\usepackage{scalerel}

\DeclareRobustCommand{\VAN}[3]{#2}
\let\VANthebibliography\thebibliography
\def\thebibliography{\DeclareRobustCommand{\VAN}[3]{##3}\VANthebibliography}

\usepackage{booktabs}

\usepackage{graphicx}	% Including figure files
\usepackage{amsmath}	% Advanced maths commands
\usepackage{amsmath}

\title[Asymmetric Errors in Transmission Spectra]{Investigating the Influence of Asymmetric Errors on Retrievals of Exoplanet Transmission Spectra}

\author[J. J. Davey et al.]{
Jack J. Davey,$^{1}$\thanks{E-mail: jack.davey.22@ucl.ac.uk (JJD)}
Kai Hou Yip,$^{1}$
Quentin Changeat$^{2}$
and Ingo P. Waldmann$^{1}$\\
$^{1}$Department of Physics and Astronomy, University College London, Gower Street, WC1E 6BT London, UK\\
$^{2}$Kapteyn Institute, University of Groningen, 9747 AD Groningen, NL.
}

\date{Accepted XXX. Received YYY; in original form ZZZ}

\pubyear{2024}

\begin{document}
\label{firstpage}
\pagerange{\pageref{firstpage}--\pageref{lastpage}}
\maketitle

% Abstract of the paper
\begin{abstract}
In studies of exoplanet atmospheres using transmission spectroscopy, Bayesian retrievals are the most popular form of analysis. In these procedures it is common to adopt a Gaussian likelihood. However, this implicitly assumes that the upper and lower error bars on the spectral points are equal. With recent observations from the James Webb Space Telescope (JWST) offering higher quality of data, it is worth revisiting this assumption to understand the impact that an asymmetry between the error bars may have on retrieved parameters. In this study, we challenge the approximation by comparing retrievals using a symmetric, Gaussian likelihood, and an asymmetric, split normal likelihood. We find that the influence of this assumption is minimal at the scales of asymmetry observed in JWST observations of WASP-39 b (with a maximum asymmetry of 77\%) but we show that it would become critical with greater levels of asymmetry (e.g. an average asymmetry of 80\%). Furthermore, we stress the importance of the shape of the asymmetric distribution and the difficulty in fitting this distribution from three summary statistics (the median and an upper and lower bound on the transit depth). An asymmetric likelihood sampler will incorrectly predict parameters if the shape of the likelihood does not match that of the underlying noise distribution even when the levels of asymmetry are equal in both. Overall, we find that it is safe to use the Gaussian likelihood assumption for current datasets but it is worth considering the potential bias if greater asymmetries are observed. 
\end{abstract}

% Select between one and six entries from the list of approved keywords.
% Don't make up new ones.
\begin{keywords}
techniques: spectroscopic -- methods: statistical -- planets and satellites: atmospheres
\end{keywords}

%%%%%%%%%%%%%%%%%%%%%%%%%%%%%%%%%%%%%%%%%%%%%%%%%%

%%%%%%%%%%%%%%%%% BODY OF PAPER %%%%%%%%%%%%%%%%%%

\section{Introduction}

In studies of exoplanet atmospheres the most popular tool for data analysis is an atmospheric retrieval, a sampling and inference pipeline based on Bayesian statistics \citep{FirstRetrievals,RetrievalSummaryChap}. While previous efforts employing forward model fitting allowed the identification of prominent features in transmission spectra \citep{Barman2007_H2O, Tinetti2007_H2O,Swain2008_CH4}, as the field has developed, retrievals have become widely accepted as the state-of-the-art method since they can predict values for atmospheric parameters while simultaneously providing statistical estimates on the confidence of predictions.

This is a particularly popular method when working with transit data from space-based observatories such as the Hubble Space Telescope (HST), the James Webb Space Telescope \citep[JWST][]{Greene2016_JWSTTransmissionSpecOverview} or the upcoming Ariel mission \citep{Tinetti2018_ArielOverview}. These observations offer low to mid-resolution spectra and, in such cases, the retrieval analyses the spectrum and can infer thermal, chemical, and physical properties of the planet such as the planet's temperature and radius, the abundances of molecules in the atmosphere, and the presence (or absence) of clouds \citep{Kreidberg2014_CloudsSuperEarth,MacDonald2017_PatchyClouds_HD209,Barstow2020_Clouds,Gao2021_CloudsReview,Ma2023_YunMaClouds,Arfaux2023_CloudsAndHaze_WASP39b}. By combining analysis of many exoplanets, we aim to better understand the diversity in the known population \citep{Sing2016_TenHotJupiters,Barstow2017_TenHotJupRetrieval,Fisher2018_HSTPopStudy,Pinhas2018_StellarContamination,Tsiaras2018_HSTPopStudy,changeat2022_PopStudy,Edwards2023_HSTPopStudy} and, as a result, to improve our understanding of the mechanisms by which planets form \citep{Oberg2011_IceLines,Zhu2021_ExoStatisticsReview,Pacetti2022_DiskChemDiversity}. However, this goal remains dependent on accurate inference for individual planets first.

Atmospheric transmission spectra are produced by fitting lightcurves to each spectroscopic channel (or binned samples thereof) from a transit observation \citep{Mandel2002_LightcurveEqns,Gimenez2006_LightqcurveEquations,Kreidberg2015_Batman,Yip2020_LightcurveRetrieval,Morvan2021_PyLightCurve-Torch}. These model the changing flux from the system when the planet is out of transit and when it occults part of the stellar surface. All lightcurves have the same general shape but they are sensitive to parameters such as the assumed limb darkening of the star \citep{Abubekerov2013_LimbDarkeningLaws} and the orientation of the system, instrumental effects, as well as the features of interest, namely, the planetary and atmospheric parameters.

Similar to the final atmospheric retrievals, sampling of lightcurve parameters employs techniques such as Markov Chain Monte Carlo (MCMC) methods to sample the depth of the transit (the difference between the in and out of transit flux) for each wavelength of observation. It is these measurements that comprise the final spectrum and the shape of the posterior distributions on these parameters which inform the errors on the measurements. Thus, it is important to keep in mind that the transit depth and its associated errors are simply summary statistics of the lightcurves.

Ideally, we would fit the atmospheric parameters directly from the lightcurves. Doing so, we would avoid any holes in the pipeline between these two analysis steps and we could better account for the covariance between the lightcurve channels. The influence of such correlated noise in retrievals as has already been investigated by \cite{Ih2021_CorrelatedNoise}. These authors find that there is a bias introduced by neglecting these effects and, in particular, stress the importance of considering these effects when fitting cloud properties with JWST data. The difficulty in fitting cloudy spectra is further highlighted by \cite{Constantinou_JWSTRetrievals_MiniNeptunes} and \cite{Davey2024_SpecBinning}. Attempting to fit atmospheric parameters directly from the lightcurves has also been attempted in previous studies \citep{Yip2020_LightcurveRetrieval,Changeat2024_LightcurveRetrieval}. However, this was found to rapidly increase the dimensionality of the parameter space under investigation and, as such, retrievals become computationally demanding. The result of this is that most studies continue to rely on a two-step process where we first fit the lightcurves and, subsequently, fit the spectrum.

Until recently, the datasets available to the exoplanet community were limited in their quantity, resolution and wavelength coverage. State-of-the-art spectra were observed with HST's Wide Field Camera 3 (WFC3) which covered the wavelength range of 0.8\,\textmu m - 1.6\,\textmu m at a spectral resolution of R\,$\approx$\,64 \citep{Dressel2023_HSTWFC3Hndbk}. With JWST, we now have access to space-based data of a greater resolution than before. Combining each of the available instruments, these data span 0.6 \textmu m - 12 \textmu m covering much of the near- and mid-infrared. Ariel will continue to supplement this data by observing around 1000 exoplanets once it launches in 2029 and, as such, this is the time to ensure that we can trust our analysis techniques in an age of improved data quality.

In the process of running an atmospheric retrieval, we have to make several assumptions and approximations for the sake of computational efficiency. These may include simplifications of the physical models such as assuming the homogeneity of the stellar surface \citep{Pinhas2018_StellarContamination,Thompson2024_StellarContamination}, a lack of variation between the day and night terminators \citep{Caldas2019_DayNightInhomog,Lacy2020_DayNightInhomog}, the use of an isothermal temperature profile \citep{Taylor2020_BiasEmissionSpecModel,Lueber2024_InfoContentJWSTWASP39b,Schleich2024_TpProfileComplexity}, or the use of free chemistry rather than equilibrium or disequilibrium chemistry models \citep{Changeat2019_TwoLayerChem,Al-Refaie2022_CompareChemModels}. However, these assumptions and approximations extend beyond our physical models to the sampling procedure itself. Both classical, Bayesian sampling algorithms and machine learning methods \citep{Nixon2020_ML4Retrievals,Yip2021_DeepLearningModels,Gebhard2022_FlowMatchRet,Vasist2023_NPERetrieval,ArdevolMartinez2024_FlopITy,Lueber2025_FASTER_LFIRetrievals} are used in the field but here we focus on the former.

In an atmospheric retrieval, we develop a posterior distribution for a set of parameters $\bar{\theta}$ (describing the atmosphere), conditional on observed data $X$ (which in our case are the observed transit depths). We require a pre-defined prior distribution $P (\bar{\theta)}$ (often a uniform distribution when nothing is known about the parameter a priori). From the parameters $\bar{\theta}$ we generate a forward model and, from this, a likelihood $P (X \vert \bar\theta )$ of the observed data $X$ given $\bar{\theta}$ can be evaluated. Bayes' theorem then gives the posterior density
\begin{equation}
    P (\bar\theta \vert X ) = \frac{P (\bar\theta) \cdot P (X \vert \bar\theta )}{P (X)}.
\end{equation}

Several studies have looked at user biases in this framework. Considering the right-hand side of the equation, the choice of prior distribution has been investigated \citep[e.g.][]{Line2013_Prior} with both uniform and more informative distributions. The evidence, $P (X)$, serves to normalise over the entire parameter space and is of less importance unless considering model comparison. In this work we consider the influence of the third element, the likelihood function. Typically this is assumed to be a Gaussian and the value of the likelihood for a given point would have the same value if the data were at a given distance above or below the model spectrum.

\begin{figure*}
    \includegraphics[width=\textwidth]{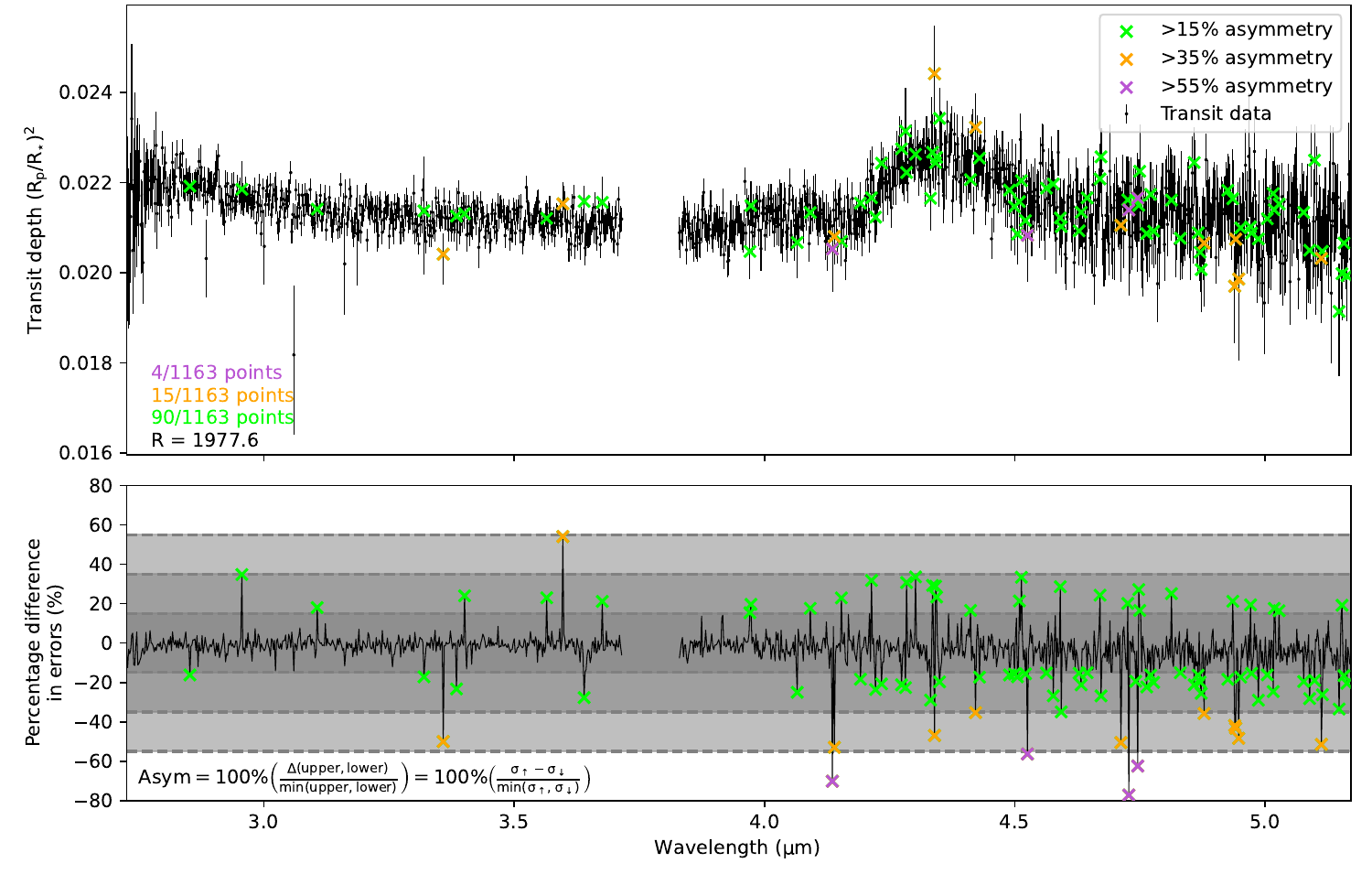}
    \caption{An observation of WASP-39 b with the JWST NIRSpec G395H instrument as published by \protect\cite{Carter2024_wasp39b}. Data cover both the NRS1 and NRS2 detectors hence the gap observed between 3.7\,\textmu m and 3.8\,\textmu m. This dataset contains different values for the upper and lower error bars on each spectral point and, in this plot, cases where the percentage asymmetry between these values exceeds 15\%, 35\% or 55\% have been indicated by green, orange and purple crosses respectively. The average spectral resolution, R, of this dataset is 1977.6.}
    \label{fig:wasp39b_G395H}
\end{figure*}

Under the current data processing framework fitting lightcurves separately to atmospheric spectra, it is imperative that we carry through as much information as possible between each step in the full data reduction and retrieval pipeline. As can be seen in the observed JWST data (an example of which is shown in figure \ref{fig:wasp39b_G395H}), in many cases, the posteriors of the transit depth are not symmetrical. When reported as a spectrum, these posterior distributions of transit depth are commonly summarised by the central value (the median of the lightcurve posterior) and the upper and lower bounds (the 84\textsuperscript{th} and 16\textsuperscript{th} percentiles of the lightcurve posterior) at each wavelength. 

The Gaussian likelihood cannot account for this asymmetry. It can only accept one value of error (the variance of the Gaussian distribution), which presents a leak in the pipeline. This could lead to less precise or, worse, incorrect predictions of parameters particularly with the improved signal-to-noise ratio (SNR) of modern data with JWST.

The sources of non-Gaussian noise in the spectra will be a combination of astrophysical and instrumental effects. Regardless of the source these present as asymmetric distributions for the lightcurve posteriors. It is beyond the scope of this paper to identify the specific sources of these asymmetries; rather, we seek to mitigate their influence given that they exist.

Several attempts have been made to incorporate simulation based inference (SBI) methods into atmospheric retrievals \citep{Vasist2023_NPERetrieval,Gebhard2022_FlowMatchRet,Lueber2025_FASTER_LFIRetrievals} and, in other fields of astrophysics, the case of non-Gaussian error distributions has also been explored \citep[e.g. velocity distributions of stars;][]{Sanders2020_AsymDists}. SBI methods bypass the need for explicit specification of the likelihood function. However, their acceptance within the field is often hampered by their reduced interpretability when compared to classical, Bayesian techniques. In some cases, the authors fit for an additional component of error in their retrievals \citep{Line2015_ErrInflation,Kitzmann2020_ErrInflationHELIOS,Lueber2025_FASTER_LFIRetrievals}. This component is designed to compensate for either an imperfect reduction of the data prior to the retrieval (i.e. an incorrect likelihood function) or incomplete physical modeling of the true scenario. We attempt to show the influence of the first of these issues explicitly and explore solutions in a Bayesian framework.

For this investigation, we use the exoplanet, WASP-39\,b \citep{Wasp39b_Discovery} as a test case since it has been observed extensively with JWST \citep{NIRCam_WASP39b,NIRSpecG395H_Wasp39b,NIRISS_WASP39b,NIRSpecPRISM_Wasp39b}. It is a hot, Saturn-sized planet orbiting a relatively inactive star. Its short orbital period and large atmospheric scale height make it an ideal target for transmission spectroscopy and, as such, it has served as a testing ground for JWST's spectroscopic instruments. These datasets were each originally analysed independently but several recent papers have attempted to study the combined dataset while considering the possible influence of different times of observation and offsets between the different instruments \citep{Lueber2024_InfoContentJWSTWASP39b,Carter2024_wasp39b}.

We base most of our analysis on the observation of WASP-39\,b using JWST's NIRSpec instrument through the G395H filter as reported by \cite{Carter2024_wasp39b}. Specifically, these data have been processed using the \textsc{ExoTIC-JEDI [V2]} reduction \citep{Alderson2022_ExoTiCJEDI} and these authors provide both an upper and lower error bar for every spectral point making it an ideal test case for this investigation. The G395H transmission spectrum from this work is shown in figure \ref{fig:wasp39b_G395H} and, on each point, the level of asymmetry is quantified and plotted in the lower panel. Asymmetry is defined to be
\begin{equation}
    \mathrm{asym} = 100\%\left(\frac{\sigma_{\uparrow}-\sigma_\downarrow}{\mathrm{min}(\sigma_{\uparrow},\sigma_{\downarrow})}\right)
\end{equation}
where $\sigma_{\uparrow}$ and $\sigma_{\downarrow}$ are the upper and lower errors respectively. There is significant variation in the observed asymmetry across the spectrum; it reaches a maximum magnitude of 77\%, which we consider to be a significant deviation from Gaussian. 

\begin{figure*}
    \includegraphics[width=\textwidth]{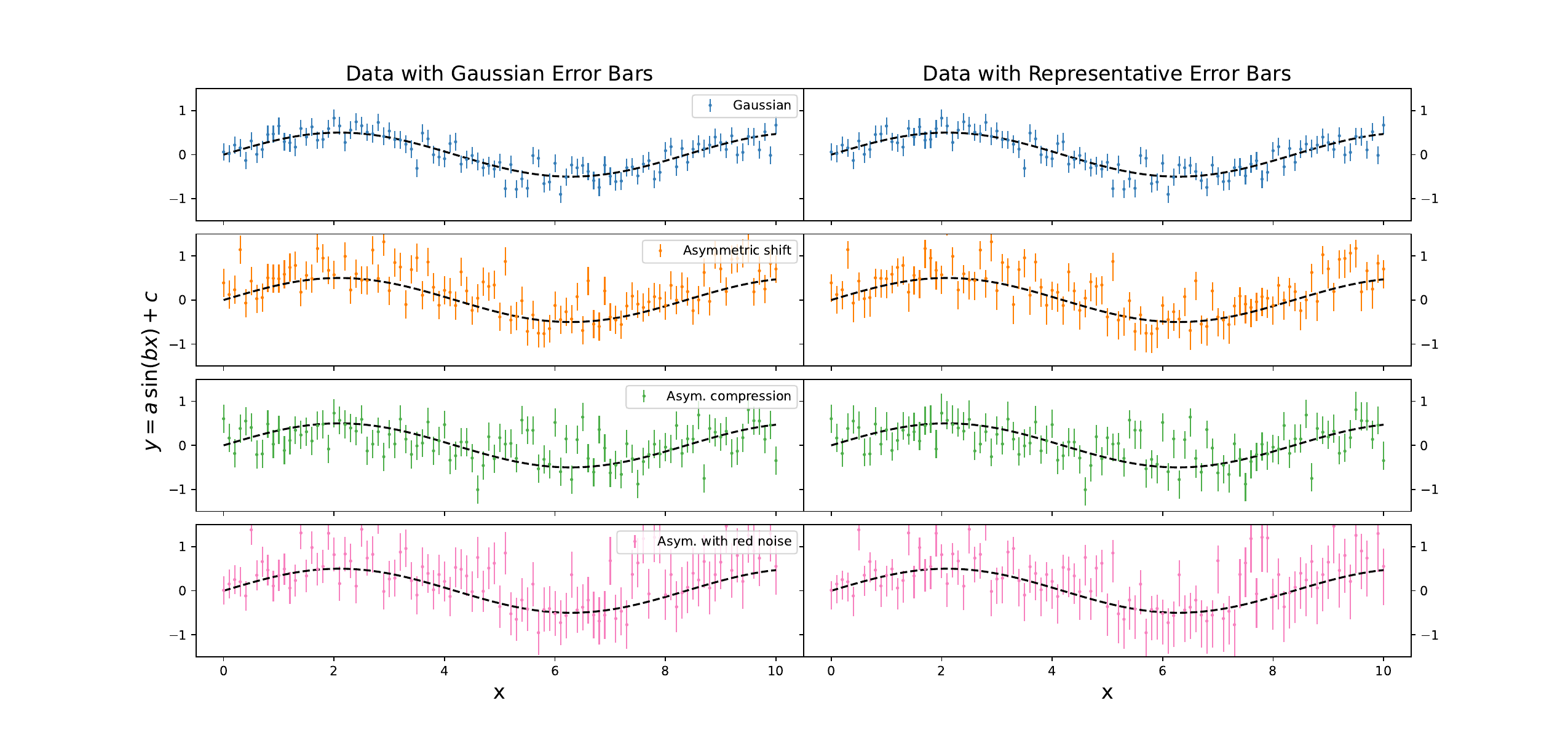}
    \caption{The simulated datasets for the four noise cases on the sine wave model. With reference to equation \ref{eq:sine}, the true values for a, b and c are 0.5,0.75 and 0 respectively. In each panel, the black dashed line displays the true, underlying model and the coloured points show the noisy data. From top to bottom, the sine wave data have noise added according to the Gaussian case, the asymmetric shift case, the asymmetric compression case and the asymmetric case with red noise scaling respectively. Each of these noise cases is described in section \ref{sec:SineWave_Methods}. In the left-hand column, the data are shown with averaged, Gaussian error bars whereas in the right-hand column, data are shown with error bars representative of the noise distribution (but reversed such that the asymmetry on the error bar is opposite to that of the noise distribution).}
    \label{fig:sine_scattered}
\end{figure*}

The extent to which the Gaussian likelihood assumption may affect the retrieved atmospheric parameters further down the pipeline is currently unknown. With the recent improvement in the quality of spectroscopic data available with regards to the achievable spectral resolution and SNR, it is appropriate to test this assumption in a rigorous manner. As such, we present a method to try to account for these asymmetries using the \textsc{TauREx3} \citep{TauREx1,TauREx2,TauREx3} retrieval code.

In this work, we present a proof of concept on a simple sine wave example, illustrating the point that sampling algorithms will retrieve different values depending on the complete or incomplete description of the noise distributions. Then, we analyse the extent of the problem in current datasets and the predicted effect as a function of the scale of asymmetry in simulated data based on the observation of WASP-39\,b. We also give consideration to cases where the asymmetric distribution used in the likelihood provides a more precise, but still incomplete, description of the true noise distribution to test the retrieval's sensitivity to the specific shape of the asymmetric distribution. Finally, we consider whether or not the influence of these asymmetries can be mitigated by higher resolution spectroscopic data. This is investigated by analysing a spectrum binned to the resolution and wavelength range of HST WFC3. 

\section{Methods}

For the purposes of this investigation we produce several simulated datasets and we use two different retrieval methods. In \ref{sec:Dists}, we provide full descriptions of the statistical distributions used. Then, in sections \ref{sec:SineWave_Methods} and \ref{sec:SimulationMethod} we detail the toy sine wave model and the parameters used in our simulations of WASP-39\,b. Section \ref{sec:ScatterMethod} outlines the non-Gaussian noise distributions used and in section \ref{sec:RetrievalMethods} we describe the two strategies used to retrieve the atmospheric parameters.

\subsection{Statistical Distributions}
\label{sec:Dists}

We will mainly be concerned with four statistical distributions. For all tests we are comparing our results against the Gaussian assumption (normal distribution) prevalent in the field. The simplest way to include asymmetry in our retrieval processes is through the approximation of a split normal distribution. However, in some cases, we use a 'custom' asymmetric noise distribution to simulate a case where there is a mismatch between the noise distribution and the likelihood used in the sampling (despite their asymmetries being equal). The fourth distribution is an exponentially modified normal distribution which we consider in cases where we attempt to fit the shape of the distribution based on only three summary statistics (a central value and an upper and lower error bar).

Details of each distribution and, where possible, probability density functions (PDFs), are provided below.
\begin{enumerate}
    \item \textbf{Normal Distribution:}
        \begin{equation}
            f_{\scaleto{\mathrm{NORM}}{3pt}}(x; \mu, \sigma) = \frac{1}{\sqrt{2\pi \sigma^{2}}}\exp\left({-\frac{(x-\mu)^{2}}{2\sigma^{2}}}\right)
        \label{eq:NormalPDF}
        \end{equation}
        where $\sigma$ is the standard deviation and $\mu$ is the mean (equivalent to the mode and the median).
        
        \,
        
    \item \textbf{Split Normal Distribution:}
        \begin{equation}
            f_{\scaleto{\mathrm{SPLIT}}{3pt}}(x; \mu, \sigma_{\uparrow}, \sigma_{\downarrow}) = A\exp\left({-\frac{(x-\mu)^{2}}{2\sigma^{2}}}\right)
        \label{eq:SplitNormalPDF}
        \end{equation}
        where the normalising constant $A = \sqrt{2/\pi}(\sigma_{\uparrow}+\sigma_{\downarrow})^{-1}$, and the standard deviation, $\sigma$, is $\sigma = \sigma_{\uparrow}$ when $x > \mu$ and $\sigma = \sigma_{\downarrow}$ when $x < \mu$. Here, $\mu$ refers to the mode of the distribution and the mean is $\mu+\sqrt{2/\pi}(\sigma_{\uparrow}-\sigma_{\downarrow})$.
        
        \,
        
    \item \textbf{Exponentially Modified Normal Distribution:}
        \begin{equation}
            f_{\scaleto{\mathrm{EXPN}}{3pt}}(x;\mu,\sigma,\lambda) = \frac{\lambda}{2} \exp\left({\frac{\lambda}{2}(2\mu+\lambda\sigma^{2}-2x)}\right) \mathrm{erfc}\left(\frac{\mu+\lambda\sigma^{2}+x}{\sqrt{2}\sigma}\right)    
        \label{eq:ExpoNormalPDF}
        \end{equation}
        where $\lambda$ is a parameter that modifies the damping effect of the exponential term. The mean is $\mu + 1/\lambda$.

        \,
        
    \item \textbf{Custom Asymmetric Noise Distribution:}
        We construct a cumulative density function (CDF) based on the reported statistics by using the piecewise cubic hermite interpolating polynomial (PCHIP) method to interpolate between the 16\textsuperscript{th}, 50\textsuperscript{th} and 84\textsuperscript{th} percentiles. This preserves the monotonicity of the CDF and ensures the first derivative (i.e. the PDF) is continuous at every point. Beyond these limits (the 16\textsuperscript{th} and 84\textsuperscript{th} percentiles), the tails of the CDF are fit using an exponentially decaying function. This means that the tails of the PDF are more heavily weighted than a standard Gaussian but helps us to emphasise the disagreement between the noise and likelihood distributions used to see if an asymmetric likelihood sampler can still offer any compensation to bias in the retrievals.
        \,
\end{enumerate}
The normal and exponentially modified normal distributions are used through their implementation in \textsc{SciPy} \citep{scipy}.

The custom distribution has no concise mathematical description. It would be preferable to use a concisely defined distribution but our custom distributions give us the freedom to fit a distribution to error bars with varying levels of asymmetry. This method relies on it being the median that is the reported central value but this is common in the field including in the \textsc{ExoTIC-JEDI [V2]} \citep{Alderson2022_ExoTiCJEDI} pipeline used by the authors of the \cite{Carter2024_wasp39b} study. In appendix \ref{app:FitAsymDist} we give consideration to the possibility of fitting well-defined distributions to our summary statistics in both the case of it being the mean and it being the median that is reported as the central value.

\subsection{Sine Wave Model}
\label{sec:SineWave_Methods}

As an initial proof of concept, we perform a retrieval on a simple sine wave model. The underlying signal is defined by three parameters, $a$, $b$ and $c$, according to the equation
\begin{equation}
    y = a\sin(bx) + c.
    \label{eq:sine}
\end{equation}
We then simulate 101 points, equally spaced between $x = 0$ and $10$, and add noise to the signal at each point according to one of four noise schemes. Note that these schemes are distinct from the previous list of distributions (but they do make use of the normal and split normal PDFs).

\begin{enumerate}
    \item [(1)]\textbf{Gaussian:} Noise is added to the signal by sampling a normal distribution of fixed variance ($\sigma = 0.2$).

    \item [(2)]\textbf{Asymmetric Shift:} The noise distribution is a split normal with a +120\% asymmetry. The lower error has a magnitude of 0.2 and the upper, 0.44, for every point. This noise case should impact on the retrieved $c$ parameter.

    \item [(3)]\textbf{Asymmetric Compression:} The noise distribution is a split normal where the asymmetry is dependent on the true y value. The errors are calculated as $\frac{3}{25} \left( \frac{8}{3}\pm \left(\frac{y}{\mathrm{max}(y)} \right)\right)$. If $y~>~0$, the sign within the brackets is negative for the upper error and positive for the lower error (and vice-versa for $y\leq0$). This results in a preferential sampling towards $y\,=\,0$ with a maximum asymmetry of 120\% (when $y~=~\mathrm{max}(y)$) and a combined error size  of 0.64 to match the asymmetric shift case. This acts to compress the wave which should impact the $a$ parameter.

    \item [(4)]\textbf{Asymmetric with Red Noise:} The noise distribution is a split normal with an upper error of $2.2\left(0.2 + \frac{x}{50}\right)$ and a lower error of $0.2 + \frac{x}{50}$. This results in a constant positive asymmetry of 120\% on every point but an error magnitude varying with $x$ to be larger at higher values (i.e. redder wavelengths in the spectroscopic case). Again, we expect this to impact the retrieved $c$ parameter.
\end{enumerate}

In each case we opt for a maximum asymmetry of +120\% to exaggerate the non-Gaussianity and to make the effects more pronounced in subsequent retrievals. Examples are shown in figure~\ref{fig:sine_scattered}. The left- and right-hand columns of this figure show the same datasets for each row with different error bars. On the right-hand side of figure \ref{fig:sine_scattered}, the error bars are representative of the noise distribution and its asymmetry. Their values  are $\sigma_{\uparrow}$ and $\sigma_{\downarrow}$ for the lower and upper error bars respectively. Note that the asymmetry is reversed in the error bars compared to the noise distribution. On the left-hand side we show the data with the approximation of a symmetric error bar. Here, the scale of the errors is set by taking the mean of the upper and lower error bars for the data on the right-hand side. 

As will be detailed in section \ref{sec:RetrievalMethods} when using the split normal distribution, it is the mode that we report as the central value. This is to preserve the perfect agreement between the error bars displayed and the subsequent sampling methods. 

\begin{figure}
    \includegraphics[width=\columnwidth]{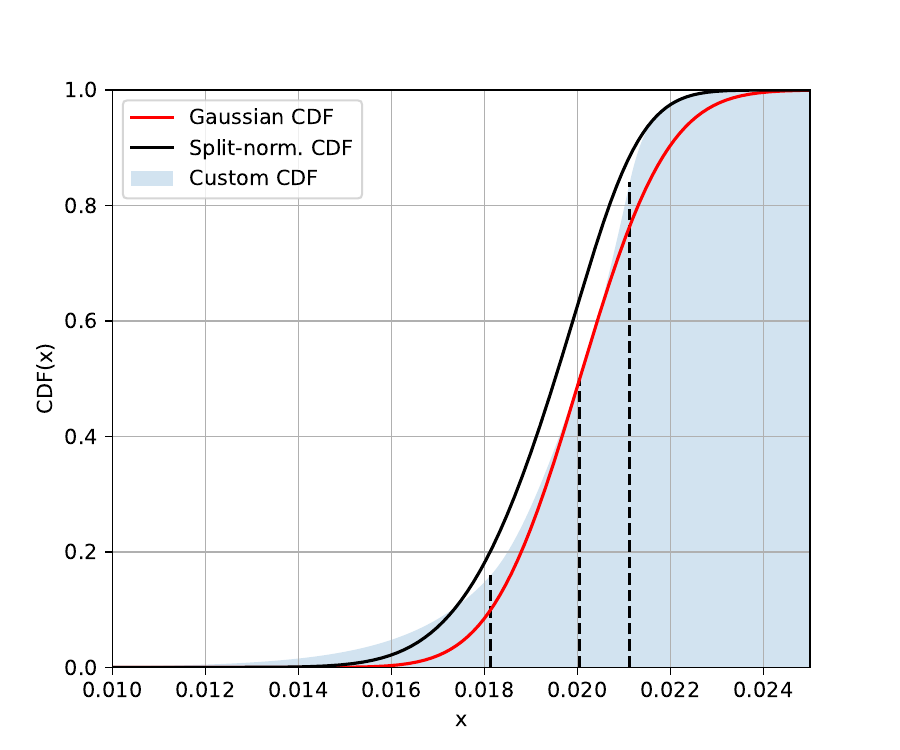}
    \caption{An example of the noise distribution and the approximation with the split normal distribution for a point in the \protect\cite{Carter2024_wasp39b} dataset for WASP-39 b observed with JWST's NIRSpec G395H instrument. We use the example of the point with the highest level of asymmetry between its upper and lower error bars (-77.21\%). The Gaussian distribution (shown in red) is parameterised by $\mu$ and $\sigma$ which are the median (set by the observed value) and the variance (set by the average of the upper and lower errors). In the case of the split normal distribution (shown in black), the widths above and below the median are set by the parameters $\sigma_{\uparrow}$ and $\sigma_{\downarrow}$ respectively. The three vertical dashed lines show the central value and the upper and lower bounds of the measured transit depth for this point. The custom CDF (used for adding noise to the data) is shown by the shaded region for comparison with the two sampling distributions.}
    \label{fig:ExampleScatter}
\end{figure}

\subsection{Simulated Transmission Spectra}
\label{sec:SimulationMethod}

We use the example of WASP-39\,b \citep{Wasp39b_Discovery} for our simulated spectra in this investigation. The spectra are designed to mimic the observations recorded by the JWST ERS teams investigating the performance and capabilities of the NIRSpec instrument \citep{NIRCam_WASP39b,NIRSpecG395H_Wasp39b,NIRISS_WASP39b,NIRSpecPRISM_Wasp39b}. The most important observation in this investigation is that taken with NIRSpec G395H since the data presented in \cite{Carter2024_wasp39b} display an average asymmetry magnitude of 5.7\% with a maximum magnitude of 77\%. This dataset is selected as it presents the highest levels of asymmetry across the JWST observations published by \cite{Carter2024_wasp39b}.

Using \textsc{TauREx3} we generate a simulation based on the parameters outlined in table \ref{table:Sim}\footnote[1]{Literature values were obtained from \url{https://exodb.space/}}. The simulation uses 100 atmospheric layers, an isothermal temperature profile, constant gas abundances throughout the height of the atmosphere, and an opaque cloud deck. This simple model serves as a convenient baseline and allows us to probe the sensitivity of our analysis methods. Additionally, we are forced to use simulations for this investigation rather than the observed dataset since we require a priori knowledge of the atmospheric inputs if we wish to comment on the accuracy of our retrieval's predictions.

We include absorption contributions from H\textsubscript{2}O, CO\textsubscript{2}, CO, SO\textsubscript{2} and Na. The remaining atmospheric content is filled by H\textsubscript{2} and He gases in a ratio of [0.17:1] as is typical for modeling of hot, gas giant planets with a primary atmosphere. Furthermore, contributions due to collisionally induced absorption from H\textsubscript{2}-H\textsubscript{2} and H\textsubscript{2}-He pairs are included. All of the opacity sources are given their appropriate references in table \ref{table:Opac}.

\begin{table}
    \centering
    \begin{tabular}{c|c|c}%|c}
    \toprule\toprule
         & \textbf{Parameter} & \textbf{Value} \\%& \textbf{Ref}\\
         \midrule
         \textbf{Planetary}  & Mass (M\textsubscript{Jup}) & 0.28\textsuperscript{a} \\
         Values from         & Radius (R\textsubscript{Jup}) & 1.27\textsuperscript{a} \\
         \cite{Wasp39b_Discovery}\textsuperscript{a}, & Isothermal Temperature (K) & 1170.00\textsuperscript{b}\\
         \cite{Barstow2017_TenHotJupRetrieval}\textsuperscript{b}, & Impact Parameter & 0.447\textsuperscript{c} \\
         and \cite{Maciejewski2016_WASP39b_OrbitalParams}\textsuperscript{c} & Orbital Period (days) & 4.055\textsuperscript{a} \\
        \midrule
        \textbf{Stellar}    & Mass (M\textsubscript{\(\odot\)}) & 0.92 \\
        All values from     & Radius (R\textsubscript{\(\odot\)}) & 1.01 \\
        \cite{Stassun2019_WASP39_StellarParams} & Temperature (K) & 5327 \\
        \midrule
        \textbf{Chemical}              & H\textsubscript{2}O & -3.5\\
        \textbf{abundances}            & CO\textsubscript{2} & -5.5\\
        (\textit{All in log scale})    & CO & -3.5\\
                                       & SO\textsubscript{2} & -6\\
                                       & Na & -5\\
        \midrule
        \textbf{Fill gases}            & H\textsubscript{2} & 0.17\\
        (\textit{Ratio of remaining}   & He & 1\\
        \textit{atmosphere}) \\
        \midrule
        \textbf{Atmospheric layers} & Cloud deck (Pa) & 1000\\
                                    & Min. pressure level (Pa) & 10\textsuperscript{-1}\\
                                    & Max. pressure level (Pa) & 10\textsuperscript{6}\\
        
    \end{tabular}
    \caption{The parameters used to simulate a transmission spectrum for WASP-39\,b similar to that published by \protect\cite{Carter2024_wasp39b}. Literature values were used for all parameters except for the molecular abundances and cloud deck which were adjusted to mimic the observation. References are indicated by superscripts next to their values where appropriate. The set up of the simulation is derived from \protect\cite{Davey2024_SpecBinning}.}
    \label{table:Sim}
\end{table}

\begin{table}
    \centering
    \begin{tabular}{c|c|c}
    \toprule\toprule
         & \textbf{Name} & \textbf{Opacity reference} \\
         \midrule
         \textbf{Atomic and molecular} & H\textsubscript{2}O & \cite{H2O}\\
         \textbf{cross-sections} & CO\textsubscript{2} & \cite{CO2}\\
                                 & CO & \cite{CO}\\
                                 & SO\textsubscript{2} & \cite{SO2}\\
                                 & Na & \cite{Na_K}\\
         \midrule
         \textbf{Collisionaly induced} & H\textsubscript{2}-H\textsubscript{2} & \cite{H2H2_CIA_HITRAN_Abel2011}\\
         \textbf{absorption}           & H\textsubscript{2}-He & \cite{H2He_CIA_HITRAN_Abel2012}\\
    \end{tabular}
    \caption{A list of opacity sources used for producing forward models in \textsc{TauREx3}. All sources come from the ExoMol project \protect\citep{ExoMol1,ExoMol2,ExoMol3} or the HITRAN database \protect\citep{RICHARD2012_HITRAN_CIA,CIA_Hitran_2013,Gordon2022_HITRAN}.}
    \label{table:Opac}
\end{table}

These simulations are binned to the wavelength grid extracted from the \cite{Carter2024_wasp39b} dataset for NIRSpec G395H. Since several studies have looked at the trade-off between the scale of error bars and spectral resolution through binning \citep{Greene2016_JWSTTransmissionSpecOverview,Guzmán-Mesa_JWST_InformationContent_WarmNep,Constantinou_JWSTRetrievals_MiniNeptunes,Taylor2023_NIRISS_WASP96b,Davey2024_SpecBinning}, we also bin this simulation to the wavelength grid of HST WFC3 to see if the influence of asymmetric error bars is altered in a lower resolution regime. For these cases, we extract the error bars from \cite{HST_WFC3_WASP39b} but scale them to provide a constant +77\% asymmetry to the data before retrieval. 

\subsection{Photometric Errors on Simulated Spectra}
\label{sec:ScatterMethod}

The simulations receive an artificial noise component designed to mimic instrumental noise in different asymmetric scenarios. In cases where we wish to simulate the assumed Gaussian noise, we take the mean of the upper and lower errors reported by \cite{Carter2024_wasp39b} for each data point and randomly sample from a normal distribution centered on the data point and with the variance, $\sigma$, set by this mean error.

In order to reproduce the same levels of asymmetric noise seen in the \cite{Carter2024_wasp39b} data with our custom distributions, each spectral point has noise added by uniformly sampling from 0 to 1 and finding the corresponding value in the custom CDF generated through the methods described in section \ref{sec:Dists}. In our investigations of extreme noise cases, we apply a +125\% asymmetry to the errors on every data point by fixing the lower error bar's position and scaling the upper to achieve this asymmetry. Then, we generate a new custom CDF with the appropriate upper value.

In several cases, we add noise using a split normal distribution since this is also the distribution that we will be using as our likelihood in our asymmetric fitting procedure. \textsc{SciPy} has no built-in sampler for this distribution so we simulate the process in two steps. First, we randomly select whether to add noise to the data point in the positive or negative direction, weighting each possibility by the ratio of the size of the respective error bars. Then, we sample from a normal distribution centered on the data point and with a width of $\mathrm{\sigma_{\uparrow}}$ if we are adding noise in the positive direction or $\mathrm{\sigma_{\downarrow}}$ if we are adding noise in the negative. Finally, we ensure the noise on the data point has the correct sign. When we add noise according to the split normal, the asymmetry is measured with respect to the mode of the distribution so that it aligns perfectly with our likelihood sampler (described in the following section).

We also consider a compression noise case similar to that used on the sine wave except that, when considering the transmission spectrum simulations, we add noise preferentially towards the minimum point in the spectrum rather than $y=0$ as in the sine wave example. The asymmetry on each point in this example is calculated as $\mathrm{asym}_{i} = -50\left (\frac{X_{i}-\mathrm{min}(\bar{X})}{\mathrm{max}(\bar{X})-\mathrm{min}(\bar{X})}\right )$ where $\bar{X}$ is the vector of transit depth measurements.

In several instances we also apply a scaling factor to the error bars to investigate the effects of the scale of the error bar as well as the level of asymmetry. For our extreme cases we use an error bar scaling factor of 1.5. 

\begin{figure*}
    \includegraphics[width=\textwidth]{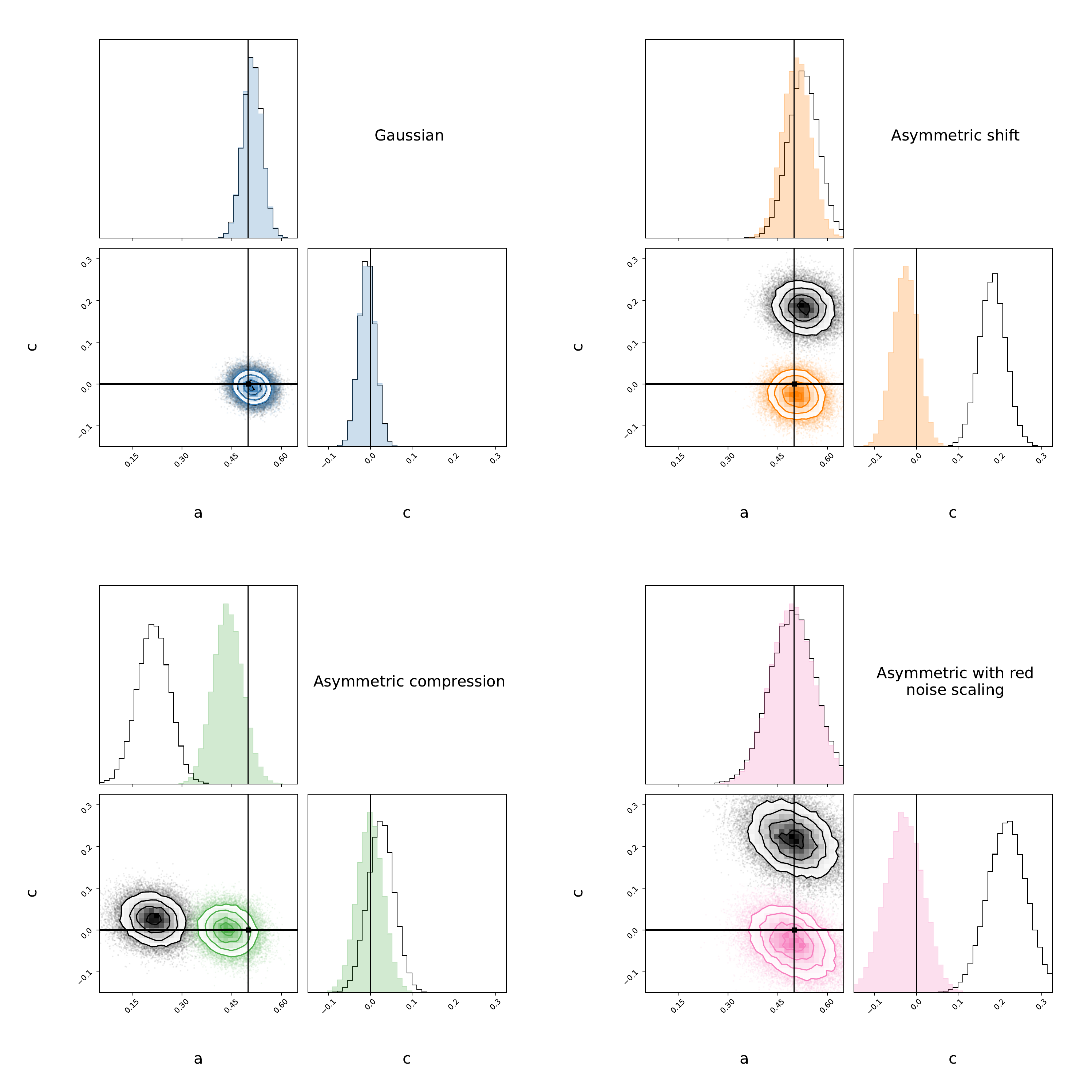}
    \caption{Comparative corner plots for each of the noisy datasets for the sine wave example. In each corner plot, true values are indicated by the horizontal and vertical black lines. The solid outlines in black display the results of the retrieval with the Gaussian likelihood and the filled histograms display the retrieval on the same dataset with the asymmetric (split normal) likelihood. In the top right of each corner plot, the underlying noise distribution for that dataset is indicated. Details of the noise schemes can be found in section \ref{sec:SineWave_Methods}.}
    \label{fig:CompareCorner}
\end{figure*}

\subsection{Retrieval Strategy}
\label{sec:RetrievalMethods}

We use an MCMC sampling procedure through its implementation in the \textsc{emcee} package \citep{emcee} to fit our data. For the sine wave tests we run 100 walkers with 20000 steps to sample the parameter space. When we analyse our outputs, we discard the first 1500 steps as burn-in.

\begin{figure*}
    \includegraphics[width=\textwidth]{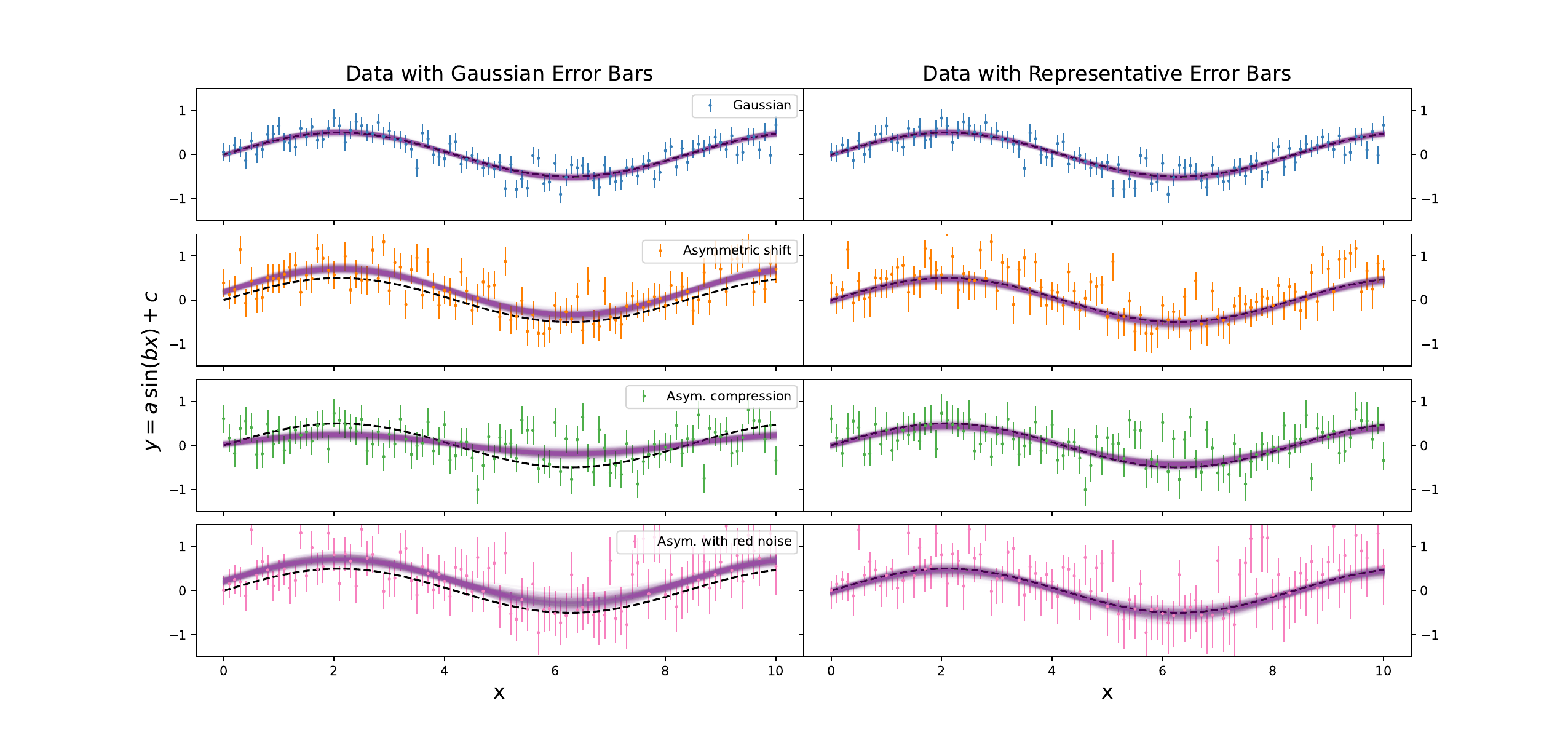}
    \caption{The simulated datasets for the four noise schemes for the sine wave model. In each panel, the black dashed line displays the true, underlying model, the purple lines show a random sample of 500 of the output solutions and the coloured points show the noisy data. From top to bottom, the sine wave data have noise added according to the Gaussian case, the asymmetric shift case, the asymmetric compression case and the asymmetric case with red noise scaling respectively. Each of these noise cases are described in section \ref{sec:SineWave_Methods}. In the left-hand column, the data are shown with averaged, Gaussian error bars whereas in the right-hand column, data are shown with error bars representative of the noise distribution (but reversed such that the asymmetry on the error bar is opposite to that of the noise distribution).}
    \label{fig:compare_sine_wave_outputs}
\end{figure*}

We run two types of retrievals on the noisy sine wave datasets. The first uses a standard Gaussian log-likelihood such that
\begin{equation}
    \ln{\left(\mathcal{L}(\bar{X})\right)} = -\frac{1}{2}\sum^{N}_{i=1}\left(\frac{(X_{i}-Y_{i})^{2}}{\sigma_{i}^{2}}\right)-\log(\sqrt{2\pi}\sigma_{i})
\end{equation}
where the $X_{i}$ and $Y_{i}$ refer to points in the model and observed spectra and the $\sigma_i$ refer to the width of the likelihood (or the scale of the error bar) on each point. 

The second retrieval uses an asymmetric likelihood as described by the split normal distribution. The form of the log-likelihood is largely the same but, in this case, for each data point we determine whether the data lie above or below the model being tested and then add a contribution to our sum with the appropriate $\sigma$ value based on the outcome of this test ($\sigma_{\downarrow}$ if the data sits above the model and $\sigma_{\uparrow}$ otherwise).

For the sine wave, we only retrieve the parameters $a$ and $c$ since we expect these to be influenced by the noise distributions outlined in section \ref{sec:SineWave_Methods} whereas $b$ controls the period of the sine wave. Uniform priors are used, between 0.05 and 4 for the $a$ parameter and between -2 and 2 for the $c$ parameter.

For retrievals of spectra, we also run an MCMC sampler through \textsc{emcee}. We use \textsc{TauREx3} as a forward model generator in these cases and sample with the same likelihoods as in the sine wave example. In some cases, we only fit for the planetary radius (R\textsubscript{p}) and the isothermal temperature (T\textsubscript{Iso}) of the atmosphere. In general, these parameters have a consistent effect across all wavelengths so we expect them to be more sensitive to our noise cases which result in a net shift up or down in transit depth. However, in order to test the sensitivity of parameters with wavelength dependent features, we also fit for the log-abundances of H\textsubscript{2}O and SO\textsubscript{2} in several cases. We chose to target these two molecules because H\textsubscript{2}O absorbs across wide molecular bands and is relatively easy to identify while, in contrast to this, SO\textsubscript{2} is a species which is harder to detect as it only has narrow features in the wavelength range of NIRSpec G395H. These two molecules also have the maximum and minimum abundances in our simulations. 

\section{Results}

We split our results into several case studies. In section \ref{sec:SineWave_Res} we analyse the fitting of our simple sine wave model. Then we process the results for the retrievals on transmission spectra working from the most extreme example towards the realistic case in sections \ref{sec:KnownScatteringSpecRet_Res} to \ref{sec:RealisticSpecRet_Res}. In these tests we only fit for the isothermal temperature and radius but, following this, we present results where molecular abundances are also fit in the retrieval (shown in sections \ref{sec:SpecMolAbun_Extreme_Res} to \ref{sec:SpecMolAbunRealistic_Res}). Finally, a low resolution case is presented to investigate the instrument dependence of our results in section \ref{sec:ResDependence}.

Throughout the results, we refer to noise and sampling distributions. In this work, noise refers to the perturbation applied to the simulated spectrum to add artificial noise and sampling refers to the likelihood distribution used during the MCMC fitting. In cases where we analyse spectra a consistent colour scheme is used. The colour scheme is outlined in table \ref{table:RetColours}. 

\begin{table}
    \centering
    \begin{tabular}{c|c|c}
    \toprule\toprule
         \textbf{Colour} & \textbf{Noise distribution} & \textbf{Sampling distribution} \\
         \midrule
         \textbf{Green} & Split normal & Asymmetric\\
         \textbf{Purple} & Split normal & Gaussian\\
         \midrule
         \textbf{Yellow} & Custom & Asymmetric\\
         \textbf{Blue} & Custom & Gaussian\\
         \midrule
         \textbf{Pale yellow} & Gaussian & Asymmetric\\
         \textbf{Pale blue} & Gaussian & Gaussian\\
    \end{tabular}
    \caption{A list of the colours used consistently throughout the results section when discussing retrieval on noisy transmission spectra.}
    \label{table:RetColours}
\end{table}

\subsection{Sine Wave Model}
\label{sec:SineWave_Res}

For the tests on the sine wave model we set the true values of $a$, $b$ and $c$ to be 0.5, 0.75, and 0 respectively. Here, we only fit for the $a$ and $c$ values as these are the ones we expect to be influenced by our asymmetric noise distributions. In all cases, $b$ is set to its true value.

In figure~\ref{fig:CompareCorner}, the black outline on each subplot shows the posterior distribution from the retrievals with the Gaussian likelihood and the left-hand column of figure~\ref{fig:compare_sine_wave_outputs} shows a subset of 500 of the sampled models for each noise case. As expected, both $a$ and $c$ are fit well when the noise is added according to a Gaussian distribution and retrieved via a Gaussian likelihood and we see the expected errors in $a$ or $c$ for the cases of the asymmetric compression and the asymmetric shift respectively. Furthermore, the red noise scaling case results in a shift in the $c$ parameter and a wide posterior centered on the correct value for $a$. 

Considering the output models shown in figure \ref{fig:compare_sine_wave_outputs}, we can see that, the retrieved solution fail to fit to the true solution (the black dashed line). This makes sense as our Gaussian retrieval procedure has no information to suggest that it shouldn't treat the data equally and symmetrically. The offsets in the $a$ and $c$ values yield a slight positive shift to the retrieved models in the second and fourth rows and a reduction in the amplitude of the wave in the third row compared to the true solution.

Results from the MCMC retrieval using our asymmetric, split normal distribution to define the likelihood function are shown by the coloured shaded histograms in the corner plot of figure~\ref{fig:CompareCorner} and in the outputs shown by the right-hand column of figure~\ref{fig:compare_sine_wave_outputs}.

The corner plot clearly shows better agreement with the true values when we account for the asymmetry in our noise distribution. Additionally, in figure~\ref{fig:compare_sine_wave_outputs}, models in the right-hand column fit the true solution very well. Thus, this provides proof of concept, highlighting the potential biases and motivating proceeding with analysis of more complicated, spectroscopic datasets. 

\subsection{Transmission Spectrum Model - Large asymmetries and a `known' noise distribution}
\label{sec:KnownScatteringSpecRet_Res}

Having used a simple sine wave model as proof of concept, we apply the same methodology to an exoplanet transmission spectrum. We use the simulation as described in section \ref{sec:SimulationMethod}.

\begin{figure}
    \includegraphics[width=\columnwidth]{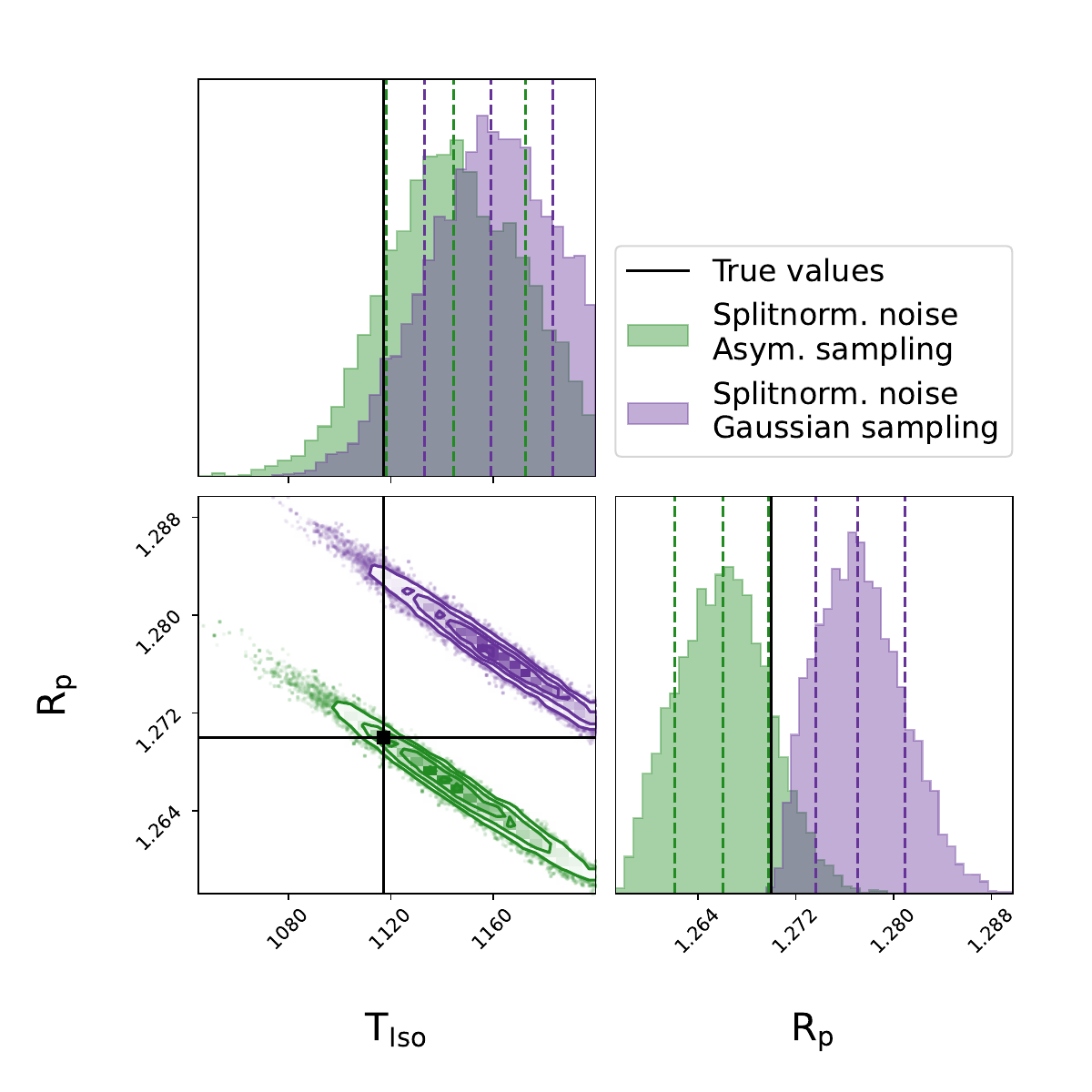}
    \caption{Retrieved results for both sampling methods when noise is added to the NIRSpec G395H WASP-39\,b simulation using to a split normal distribution with a +125\% asymmetry and where the observed error bars are scaled by a factor of 1.5. Dashed lines represent the 16\textsuperscript{th}, 50\textsuperscript{th}, and 84\textsuperscript{th} percentiles for the posterior of the corresponding colour in each histogram.}
    \label{fig:KnownScatPost_200}
\end{figure}

In the first instance we use a pessimistic asymmetric case where the asymmetry is large and, therefore, likely to influence the outputs. Here we assume that the full distribution of the noise is well known a priori. We take datasets where noise is added according to a split normal distribution with a +125\% asymmetry between the upper and lower errors and perform retrievals with both our standard Gaussian and our asymmetric likelihood samplers. Results from this test are shown in figure \ref{fig:KnownScatPost_200}.

There is a significant offset between the posteriors returned by the asymmetric sampler and the Gaussian sampler. This is somewhat evident in the one-dimensional marginals, and very clear in the full two-dimensional posterior. We see that the posterior using the asymmetric sampler includes the true value, while the posterior using the Gaussian sampler does not.

This highlights the need to look beyond the individual parameter predictions in the marginalised posteriors and to analyse the full corner plot for a better understanding of the retrieval's performance. Focusing on a single parameter could lead to false confidence in the predictor.

\subsection{Transmission Spectrum Model - Large asymmetries and `unknown' noise distribution}
\label{sec:ExtremeAsym_Res}

The previous test used the `easiest to detect' noise scenario. Now, we still consider an extreme case but we try to describe the asymmetry with an incorrect distribution.

For these retrievals, the noise is added according to a custom distribution which has a +125\% asymmetry. The sampling likelihood distribution is unchanged. We still use the split normal likelihood sampler to see if we can improve the predictions from the Gaussian likelihood despite having incorrectly described the asymmetry of the noise distribution with our likelihood. 

\begin{figure*}
    \includegraphics[width=\textwidth]{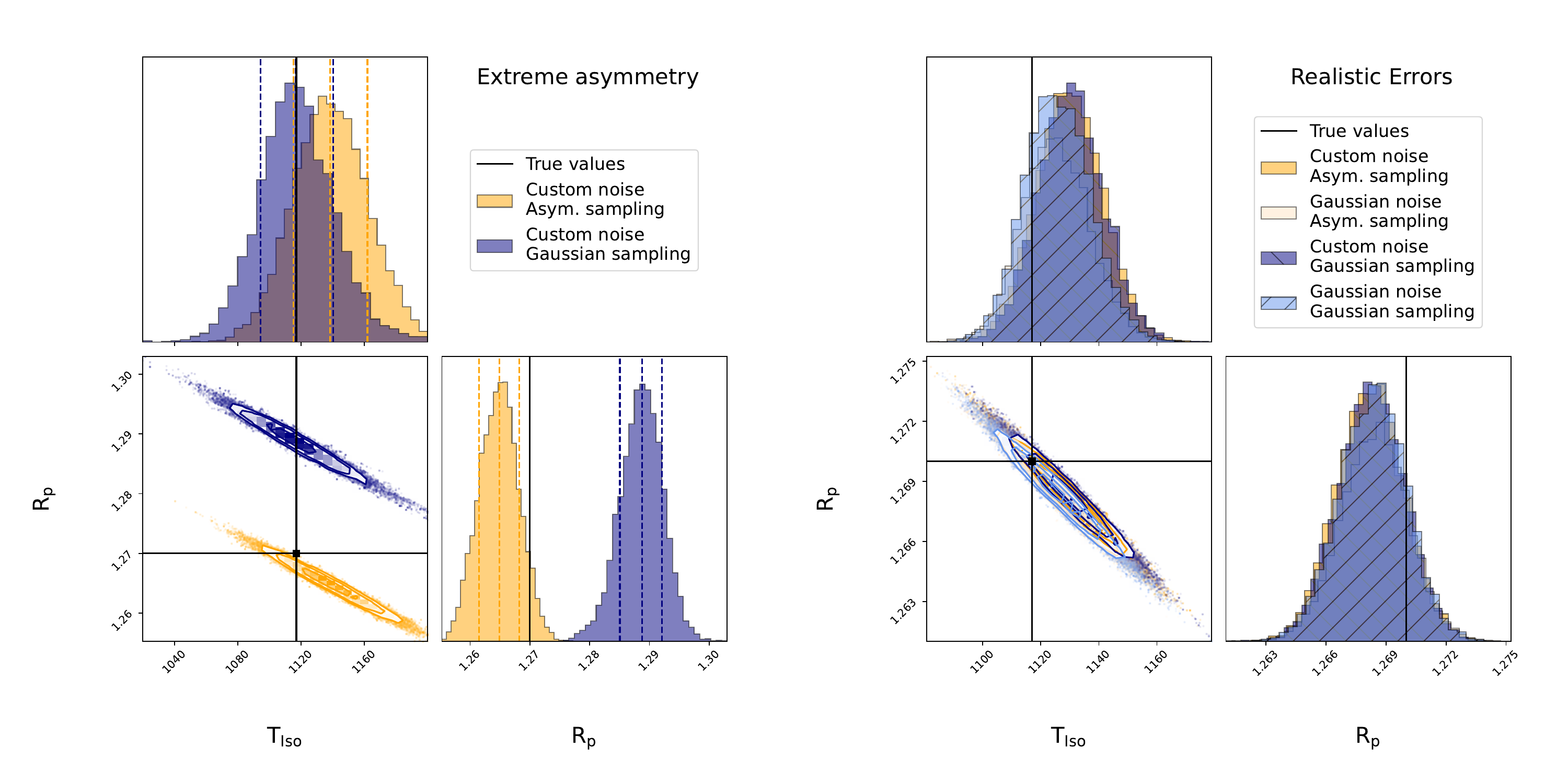}
    \caption{Retrieval results comparing the different sampling techniques on data where noise is added with either a Gaussian or a custom (asymmetric) distribution. In the right-hand corner plot, the noise is scaled to be representative of that reported in the \protect\cite{Carter2024_wasp39b} dataset whereas on the left, in the extreme case, the noise is scaled by a factor of 1.5 and a +125\% asymmetry is applied to each point. In each case, the noise describes which distribution was used to perturb the signal while the sampling describes which likelihood was used in the fitting procedure. On the left-hand corner plot, the vertical dashed lines mark the 16\textsuperscript{th}, 50\textsuperscript{th}, and 84\textsuperscript{th} percentiles of the posterior distributions. These are excluded from the right-hand corner plot for clarity.}
    \label{fig:CornerPlots_Spec}
\end{figure*}

As can be seen in the left-hand panel of figure \ref{fig:CornerPlots_Spec}, the results of these two retrievals are significantly offset from one another. Both samplers predict the true isothermal temperature temperature within the 68\% confidence interval and neither finds the correct planetary radius. However, considering the fitting of radius, the asymmetric sampler gets much closer to the true value and accounting for the asymmetry in the likelihood still makes a significant improvement. The inability to get accurate predictions on both parameters highlights the significance of our choice of asymmetric distribution in our sampling scheme and the sensitivity that the retrieval will have to this assumption.

\begin{figure*}
    \includegraphics[width=\textwidth]{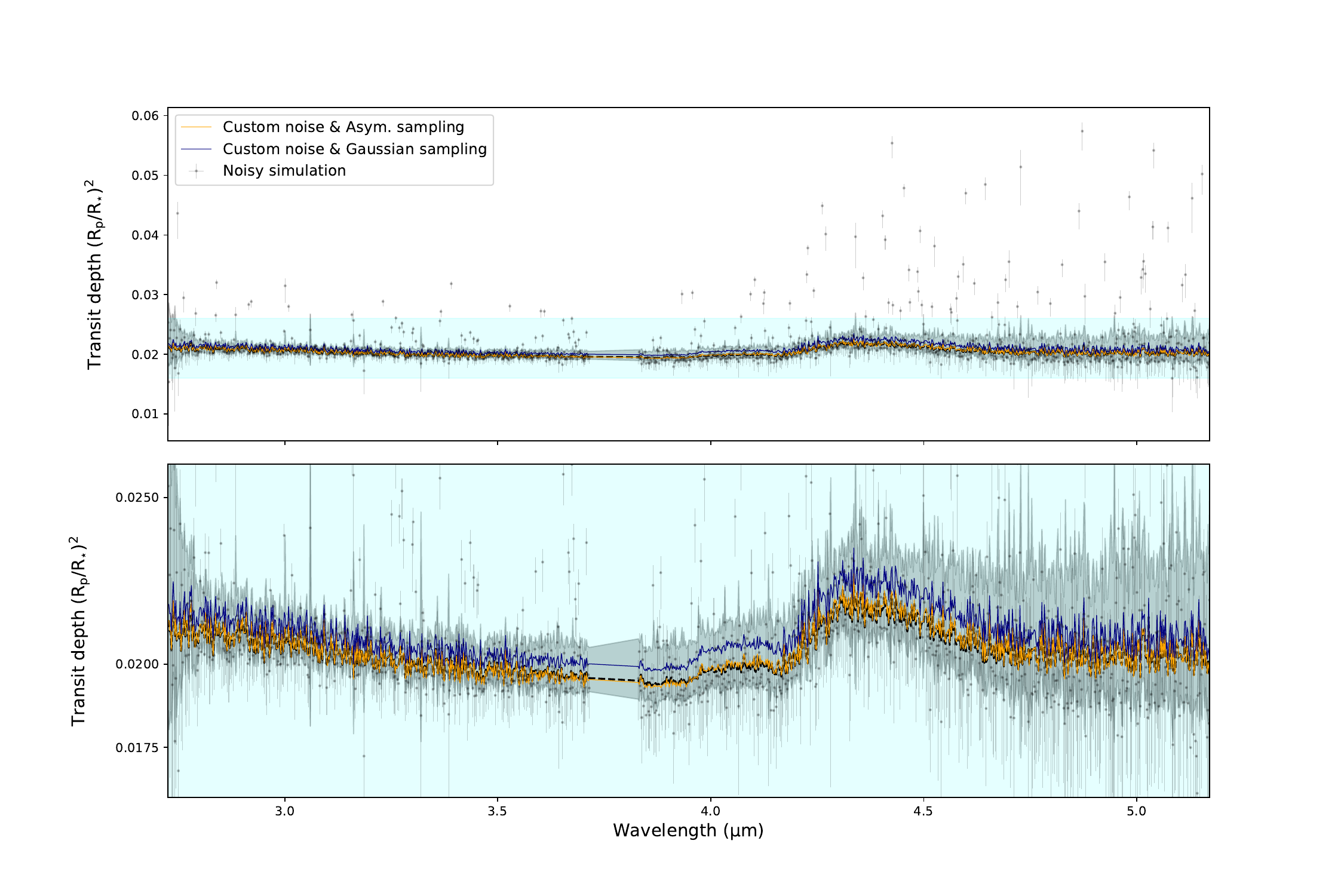}
    \caption{The retrieved spectra and input data for the case of extreme asymmetry (+125\% on each spectral point and error bars scaled by a factor of 1.5) when retrieving with a Gaussian likelihood (blue line) or an asymmetric likelihood (yellow line). The black dashed line in the upper two panels shows the true noiseless simulation while the grey shaded region shows the 1\,$\mathrm{\sigma}$ interval of the noise distributions. Given the exponential tails of the custom distribution, it is more susceptible to outliers than a Gaussian noise case. The lower panel shows a version of the data in the cyan shaded region of the upper panel for improved visibility of the differences between the retrieved solutions.} 
    \label{fig:ExtremeAsym_OutputSpec}
\end{figure*}

It is worth considering the median retrieved solutions as displayed in figure \ref{fig:ExtremeAsym_OutputSpec}. From a visual inspection and without knowing the ground truth value, it would be difficult to determine which of the sampling schemes had better estimated the true values. Despite the median retrieved values being closer in their absolute values in the asymmetric sampling case, it would be easy to believe the fitting of the Gaussian sampler based on the fit model alone. 

\subsection{Transmission Spectrum Model - Realistic asymmetries and `unknown' noise distribution}
\label{sec:RealisticSpecRet_Res}

Now we relax the second assumption of our original, pessimistic case. We no longer use `extreme' error bars and, instead, we extract errors from the observed data for WASP-39\,b. Considering the retrievals on these data, the results match for the traditional, Gaussian assumption and the asymmetric sampling case. The true value is correctly identified in each case of asymmetric noise since the sampler is not sensitive enough to the difference in the noise and the sampling distributions to make conflicting predictions. These results are shown in the right-hand panel of figure \ref{fig:CornerPlots_Spec}.

To give a quantitative measure of the agreement between the sampling methods in this realistic case, refer to the small values of the Kolmogorov-Smirnov test (KS-test) shown in table \ref{table:KS_Spectra}. This test provides a measure of the probability that our two samples are drawn from the same underlying distribution considering both their location and shape. The test statistic takes values from 0 to 1 and the closer to 0 it is, the more likely it is that the samples are drawn from the same distribution. 

\begin{table*}
    \centering
    \begin{tabular}{c|c|c|c|c|c}
    \toprule\toprule
         \textbf{Spectrum} & \textbf{Noise case} & \textbf{T\textsubscript{Iso}} & \textbf{R\textsubscript{p}} & \textbf{log(H\textsubscript{2}O)} & \textbf{log(SO\textsubscript{2})}\\
         \midrule
         \textbf{WASP-39\,b} & Split normal (+125\% asym., 1.5x scaling) & 0.225 & 0.903 & - & - \\
         \textbf{NIRSpec G395H} & Custom (+125\% asym., 1.5x scaling) & 0.364 & 0.9995 & - & -  \\
         & Custom (Realistic)  & 0.014 & 0.030 & - & -  \\
         & Gaussian (Realistic)  & 0.027 & 0.022 & - & -  \\
         \midrule
         \textbf{WASP-39\,b} & Split normal (+125\% asym., 1.5x scaling) & 0.539 & 0.977 & 0.816 & 0.158  \\
         \textbf{NIRSpec G395H} & Split normal (compression) & 0.960 & 0.508 & 0.910 & 0.361\\
         (fitting molecules) & Custom (Realistic)  & 0.066 & 0.059 & 0.018 & 0.021 \\
         \midrule
         \textbf{WASP-39\,b} & Custom (+77\%) &  0.294 & 0.179 & - & -  \\
         \textbf{HST WFC3} &  & \\
    \end{tabular}
    \caption{Results from the two-sample KS-test comparing the posterior distributions for Gaussian and asymmetric likelihood retrievals on the noisy simulations of WASP-39\,b transmission spectra.}
    \label{table:KS_Spectra}
\end{table*}

\begin{figure}
    \includegraphics[width=\columnwidth]{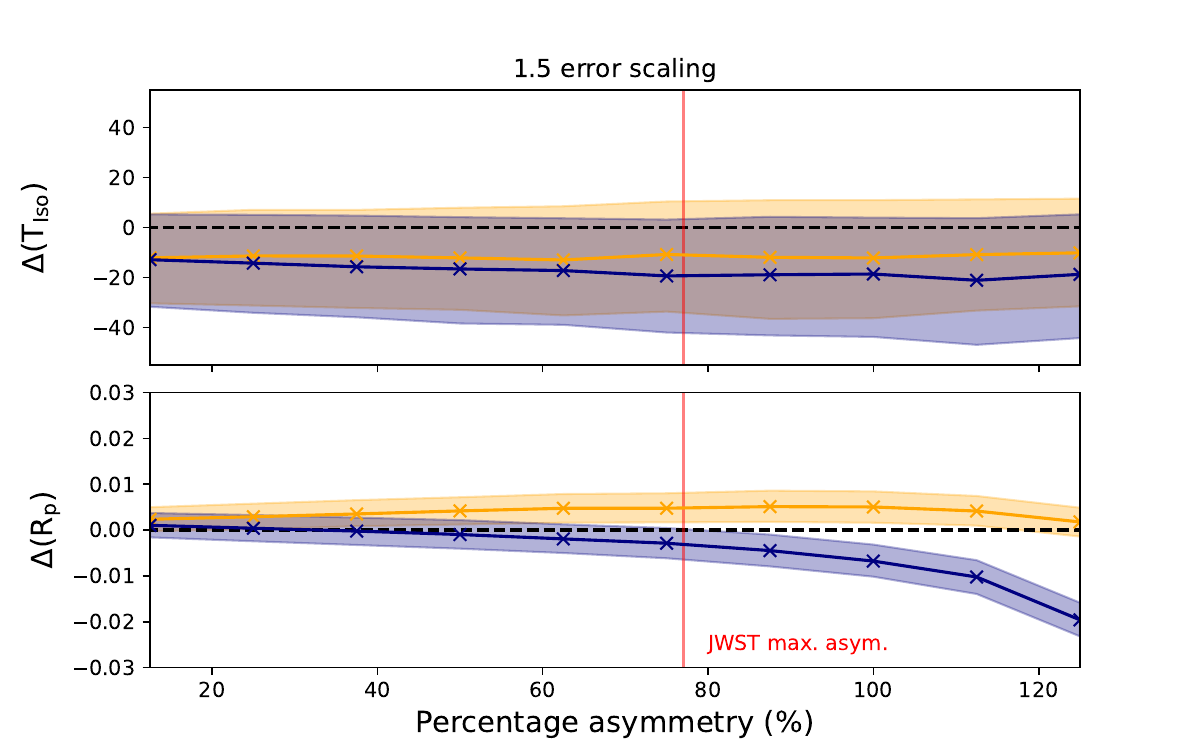}
    \caption{The deviation from the true value of the retrieved isothermal temperature and planetary radius as the level of asymmetry in the dataset is varied. These simulations of WASP-39\,b with JWST's NIRSpec G395H instrument had noise added according to a custom distribution but was retrieved with the asymmetric sampler which uses a split normal distribution. As in previous plots, the yellow and blue show results from sampling with the asymmetric and the Gaussian likelihoods respectively. In each case the solid line shows the median value and the shaded region indicates the 1$\sigma$ interval. The vertical red line marks an asymmetry of +77\% since this is the maximum magnitude of the observed asymmetries in the data reported by \protect\cite{Carter2024_wasp39b}.}
    \label{fig:CornerPlots_VaryAsym}
\end{figure}

Given the lack of sensitivity to the asymmetry in the realistic case and the clear offset in the extreme case, it is worth considering the deviation of predictions from the true value as a function of the asymmetry applied. We run both types of retrieval on data where the errors are given an enforced asymmetry between +12.5\% and +125\%. Results are shown in figure \ref{fig:CornerPlots_VaryAsym}. From these data it is clear to see that the asymmetric sampler deviates less from the true value than the predictions from the Gaussian sampler despite having misrepresented the true shape of the distribution.

Additionally, if we consider the uncertainty of our predictions (indicated by the shaded regions in figure \ref{fig:CornerPlots_VaryAsym}), they show little variation with the scale of asymmetry. This indicates that the confidence is more heavily influenced by the size of the error bar than by the asymmetry. This is a valid cause for concern. Operating with a Gaussian likelihood in an extreme asymmetry regime, we would have no indication from our results that we were misrepresenting our data and, as such, we would be overly confident in our predictions.

\subsection{Retrieving Molecular Abundances - Extreme asymmetries and `known' noise distribution}
\label{sec:SpecMolAbun_Extreme_Res}

In previous retrievals we have only fit for the isothermal temperature of the atmosphere and the planetary radius. Both of these parameters contribute to the spectrum as a near-constant offset in the observed transit depth across wavelengths. As such, we expected these to be sensitive to our noise distributions which were preferentially adding noise in the positive direction across all points. 

Since molecular abundances will present as wavelength dependent features in the spectrum, we wish to determine if these are affected differently to the radius and temperature. For these tests we retrieve the log-abundances of H\textsubscript{2}O and SO\textsubscript{2}. Of the five species included in the simulation, we choose to retrieve these two since they present extremes in our contributions. H\textsubscript{2}O is a relatively easy-to-retrieve molecule due to its broad features across the wavelength range of interest. In contrast, SO\textsubscript{2} is mainly identified by a very narrow feature at 4.1\,\textmu m. Their abundances also represent the maximum and minimum contributions in our spectrum (at log-abundances of -3.5 and -6 for H\textsubscript{2}O and SO\textsubscript{2} respectively).

\begin{figure}
    \includegraphics[width=\columnwidth]{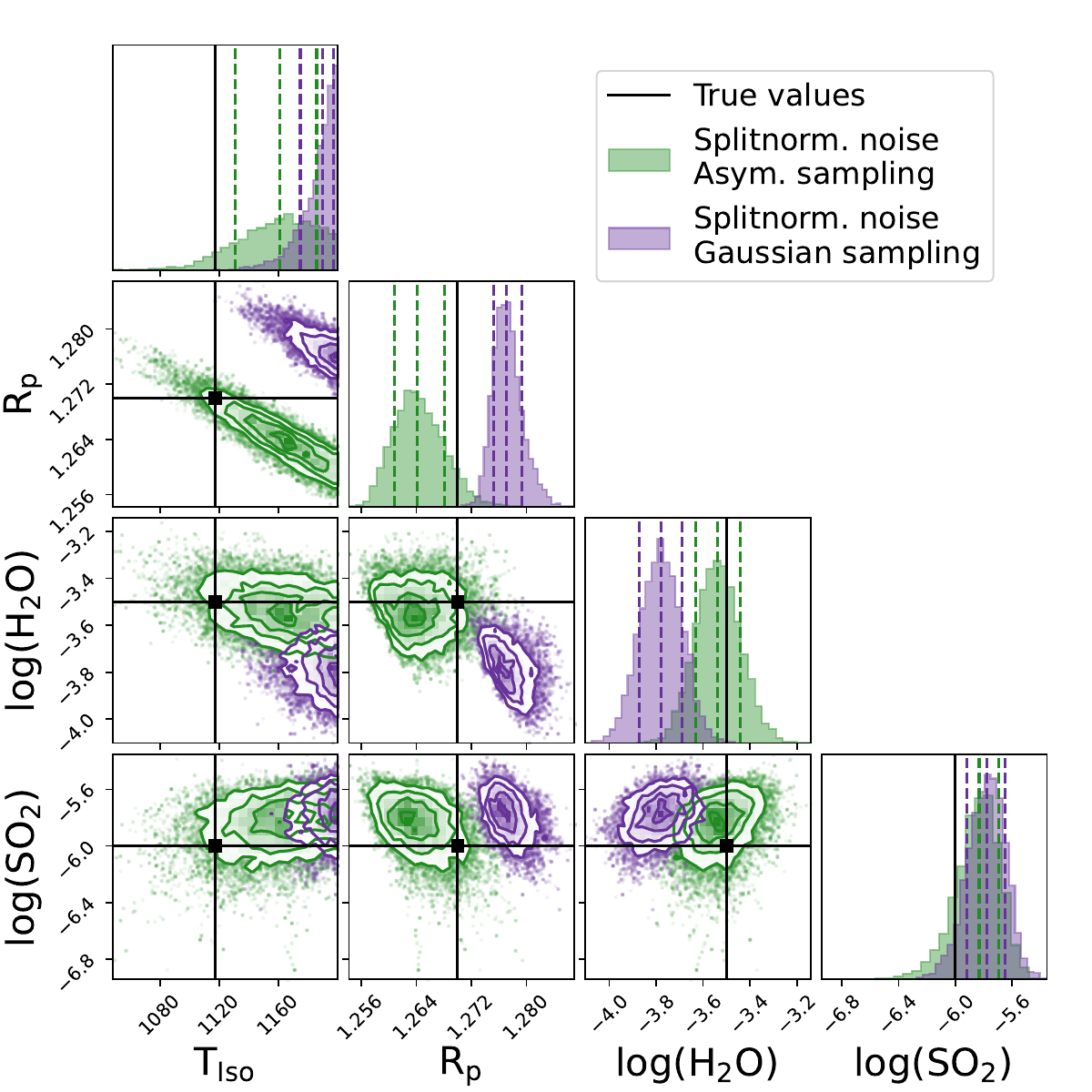}
    \caption{Retrieved results for both sampling methods when the NIRSpec G395H WASP-39\,b simulation had noise added according to a split normal distribution with a +125\% asymmetry and a scaling factor of 1.5 applied to the noise distribution on every point. Dashed lines represent the 16\textsuperscript{th}, 50\textsuperscript{th}, and 84\textsuperscript{th} percentiles for the posteriors of the corresponding colour in each histogram.}
    \label{fig:RetMol_Extreme}
\end{figure}

Figure \ref{fig:RetMol_Extreme} shows the retrieved posterior distributions when a constant +125\% asymmetry was applied to the spectrum and the errors were scaled up by a factor of 1.5 times from their observed scales. For this case, we use the split normal distribution to add noise to the data and, as such, our asymmetric sampler provides a perfect description of the error distribution. This allows us to test the sensitivity of molecular abundances to the asymmetric noise without introducing an incorrectly defined shape of the distribution as an additional source of uncertainty. 

In this case, we find our Gaussian assumption to be insufficient. In all of the parameters, it proves incorrect in its predictions, missing the true value by more than 1\,$\sigma$.

We see a vast improvement when we sample with our asymmetric likelihood distribution. It is able to correctly identify the relevant parameter space in each of the two dimensional joint posterior plots in figure \ref{fig:RetMol_Extreme} showing sensitivity to degeneracies between the parameters. As such, in this case we are sensitive to the input noise distribution in retrievals. It is not only in our wavelength independent parameters (T\textsubscript{Iso} and R\textsubscript{p}) that we observe this sensitivity. One of the parameters for which the Gaussian assumption performs the worst is the log-abundance of H\textsubscript{2}O, a broad band, but still wavelength dependent, contributor to the overall simulation.

\subsection{Retrieving Molecular Abundances - Compression case}
\label{sec:SpecMolAbun_Compress_Res}

\begin{figure}
    \includegraphics[width=\columnwidth]{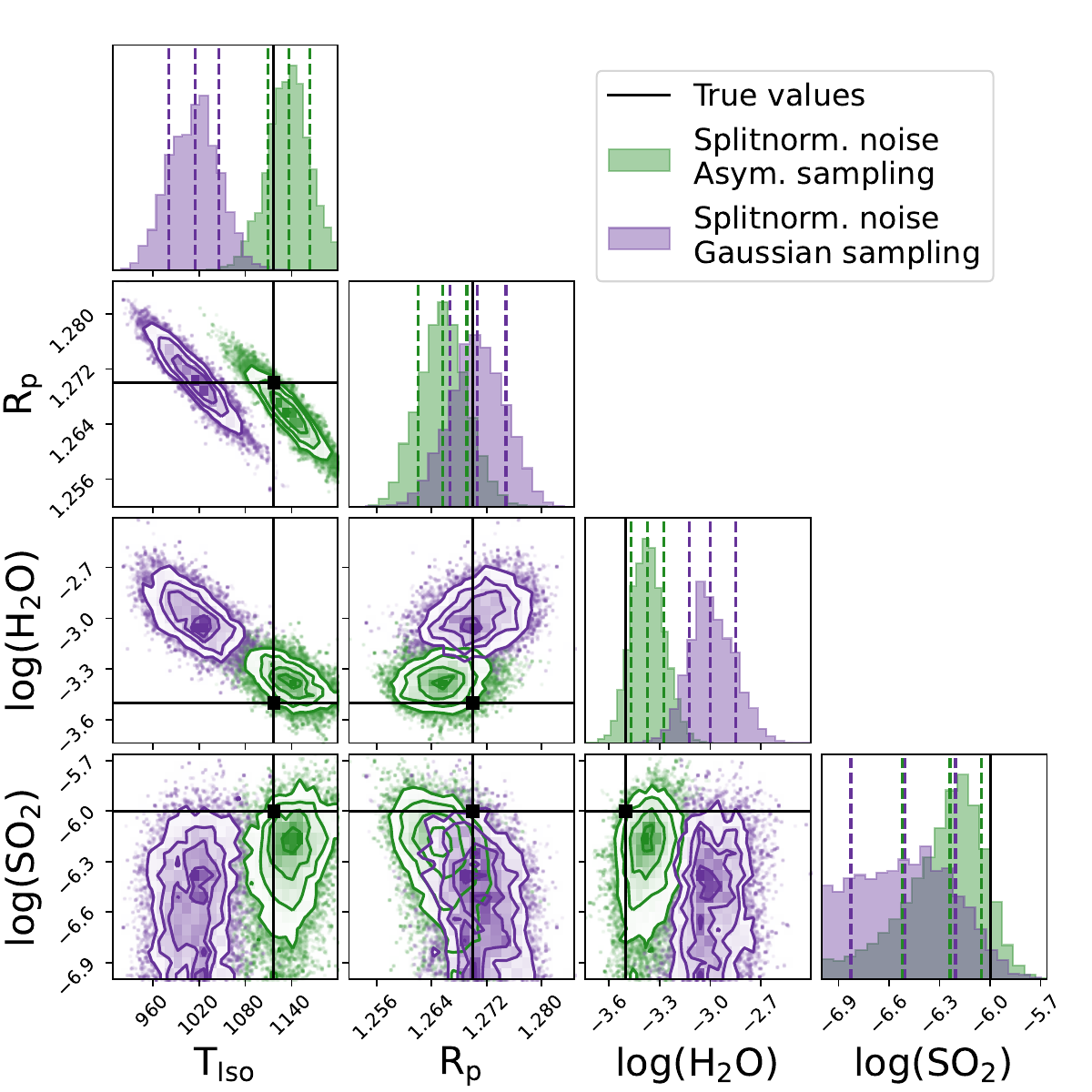}
    \caption{Retrieved results for both sampling methods when the NIRSpec G395H WASP-39\,b simulation had nosie added according to a split normal distribution designed to preferentially add noise to the signal towards the baseline of the spectrum (the compression noise case). Dashed lines represent the 16\textsuperscript{th}, 50\textsuperscript{th}, and 84\textsuperscript{th} percentiles for the posteriors of the corresponding colour in each histogram.}
    \label{fig:RetMol_Compression}
\end{figure}

Now we consider the compression noise case. These data have a $-50\%$ asymmetry at the maximum point and the asymmetry is scaled such that it reduces to 0\% at the baseline (minimum) of the spectrum. In this case, we continue to use a split normal noise case and a split normal likelihood sampler. Results from retrievals on this dataset are shown in figure \ref{fig:RetMol_Compression}.

The SO\textsubscript{2} abundance is poorly fit by both sampling methods as it has a very shallow feature in the original simulation. However, while the Gaussian likelihood sampler can only place an upper limit on the value, the asymmetric sampler offers slightly more confidence in its predictions, localising the solution around the upper edge of the Gaussian sampling's posterior. 

The Gaussian sampler retrieves a higher log-abundance of H\textsubscript{2}O compared to the asymmetric sampler. Additionally, while in previous analyses, the Gaussian sampler has retrieved a higher isothermal temperature than the asymmetric sampler, here it retrieves a lower value. The interplay between the 
wavelength independent parameters (R\textsubscript{p} and T\textsubscript{Iso}) and the molecular features means that the asymmetric sampler is able to compensate for the reduced height of features by shifting the baseline of the spectrum. If these tests are re-run while fixing the values of temperature and radius to their true inputs, as expected, the asymmetric sampler retrieves a higher abundance for both molecules and more significantly for H\textsubscript{2}O.

\subsection{Retrieving Molecular Abundances - Realistic asymmetries and `unknown' noise distribution}
\label{sec:SpecMolAbunRealistic_Res}

\begin{figure}
    \includegraphics[width=\columnwidth]{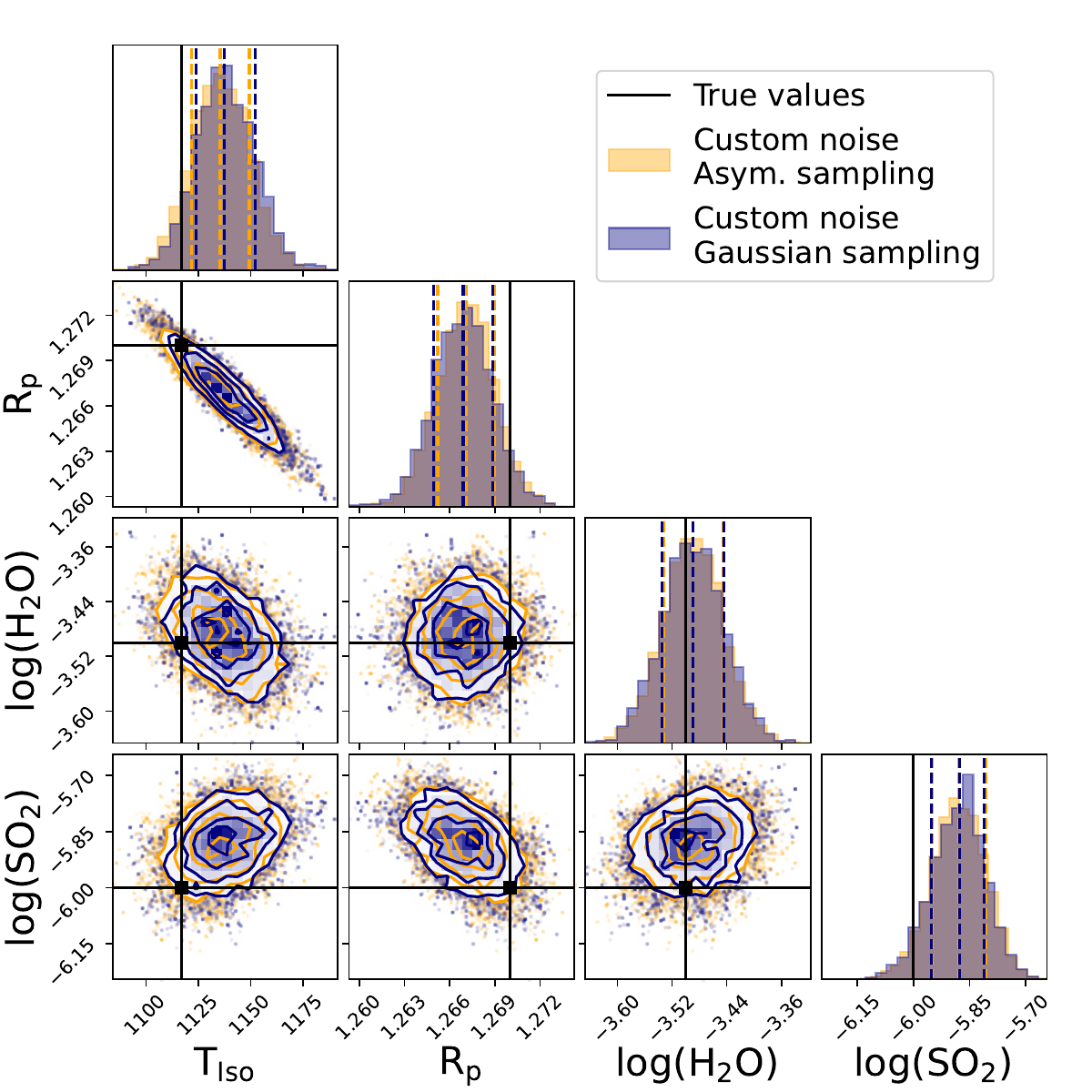}
    \caption{Retrieved results for both sampling methods when the NIRSpec G395H WASP-39\,b simulation had nosie added according to a custom distribution with realistic error bars and levels of asymmetry. Dashed lines represent the 16\textsuperscript{th}, 50\textsuperscript{th}, and 84\textsuperscript{th} percentiles for the posteriors of the corresponding colour in each histogram.}
    \label{fig:RetMol_Realistic}
\end{figure}

Finally, we consider the retrieval of molecular abundances when we have an `unknown' noise distribution (a custom distribution rather than the sampling's split normal distribution) with realistic error magnitudes and asymmetries. These results are shown in figure \ref{fig:RetMol_Realistic} and, as with the retrievals on the isothermal temperature and planetary radius alone, there is almost perfect agreement between the two retrievals in this case. This demonstrates that the Gaussian likelihood assumption is still safe when retrieving datasets with the levels of asymmetry observed in JWST data even if the retrieval is considering a greater number of parameters.

\subsection{Compensating for Asymmetry with Spectral Resolution}
\label{sec:ResDependence}

Given that our investigation is a study of the extractable information from the spectrum, we also give consideration to the resolution of our input observation. Given the previous results in this paper, we hypothesised that the bias on parameter estimations due to  asymmetric error bars may be influenced by the resolution of the input data.

For this additional investigation, we bin our simulation to the wavelength grid of HST WFC3 taken from the spectrum reported by \cite{HST_WFC3_WASP39b}. We scale the error bars on this dataset such that they have a +77\% asymmetry (similar to the maximum seen in the JWST observations) and scale the size of the error bars by a factor of 1.5 to try to emphasise the effect and remain consistent with previous analysis in this paper.

\begin{figure}
    \includegraphics[width=\columnwidth]{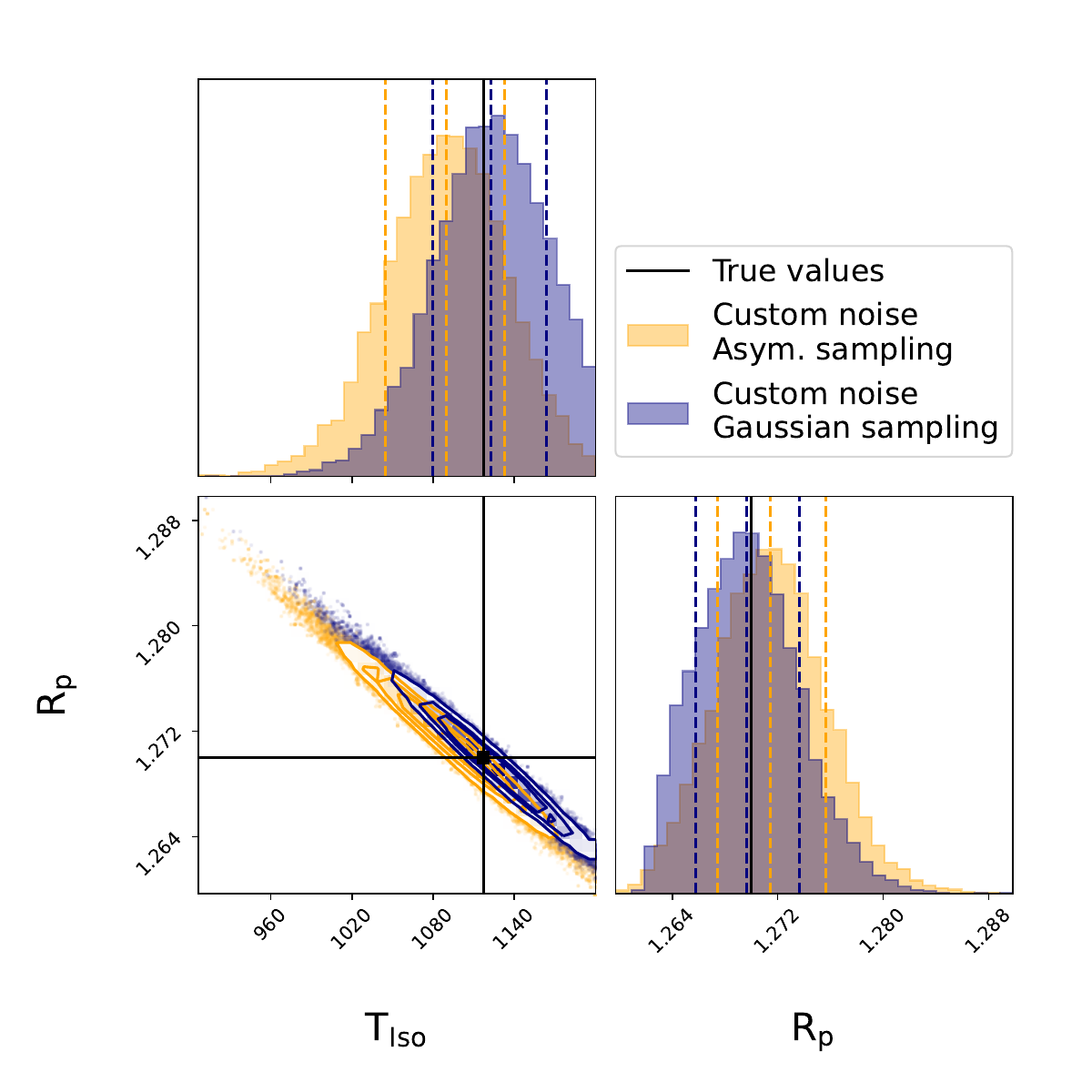}
    \caption{Retrieved results for both sampling methods when the WASP-39\,b simulation is binned to the wavelength range of HST's WFC3 instrument and noise is added according to custom distribution with a +77\% asymmetry. Dashed lines represent the 16\textsuperscript{th}, 50\textsuperscript{th}, and 84\textsuperscript{th} percentiles for the posteriors of the corresponding colour in each histogram.}
    \label{fig:HST_Corner}
\end{figure}

The results from the retrievals are shown in figure \ref{fig:HST_Corner}. We note a minor offset between our posterior plots but this is minimal and the predicted values are well within 1\,$\sigma$ of each other. Quantitative evidence of this is shown by the small value for the KS-test outlined in table \ref{table:KS_Spectra}. As such, we conclude that resolution is not a significant factor in the analysis presented in this paper and the results should be applicable to other instruments. 

\section{Discussion and Conclusion}

The purpose of this paper is not to identify the source of the asymmetry in error bars when performing atmospheric retrievals; rather, it is intended to explore the severity of failing to account for these effects. In our investigation of the sine wave model, we demonstrate how retrieved parameters can be skewed by a misrepresentation of the noise in the sampling procedure.

We investigated the use of a split normal distribution as a simple way of trying to combat the problem and approximate an asymmetric distribution of an unknown shape. This would be a practical method for a real dataset if, at best, authors provided three values for their reported data points, a central value as well as some upper and lower error bar. However, we find this to be insufficient in cases of extreme asymmetry. Here, the shape of the underlying distribution is of comparable importance to the consideration of the different upper and lower errors bars.

Therefore, we advocate for the reporting of complete posterior distributions from lightcurve fittings as only then can the information be carried through to the retrieval without additional assumptions being necessary. We investigated the possibility of fitting some arbitrarily chosen asymmetric distribution to a dataset based off of the three commonly reported values but found that our efforts were unable to resolve degeneracies in possible solutions for the underlying distribution with these summary statistics alone (see appendix \ref{app:FitAsymDist}). In addition, we would still be biased by our assumption of the shape of the distribution in these cases and can only account for this by seeing the full posterior distribution of the lightcurve fitting step.

Notably, we do not see any offset between our two retrieval methods when operating in our regime of realistic error bars both in terms of the scale of the error bar and the levels of asymmetry. It is safe to carry on under the Gaussian assumption for current datasets but it is worth considering if, in the future, we may enter our extreme asymmetry regime and would need to account for these asymmetries more carefully. 

\subsection{Limitations and Future Work}

In this work, we have only considered artificial noise cases. In many examples, these were designed with the intention of mimicking or masking the effect of a particular parameter in the simulation. While some of these do have physical interpretations (e.g. the instances of red noise), the emphasis of this study was on sensitivity to these asymmetric data, not the physical mechanisms that could lead to their existence in the first place.

It would be valuable to pursue this work further by stepping back to the lightcurve fitting stage and trying to propagate the full likelihood based on the posterior distribution of the transit depth fitting. However, we stress again that this was not the motivation of this paper. We sought only to see if we would be sensitive to these effects should they exist and given the reporting of asymmetries by authors such as \cite{Carter2024_wasp39b}. Additionally, we wanted to see how well we could compensate for their effects assuming data are presented as a central value and a pair of errors where the full distribution is unknown.

Considering the lightcurve fitting would also preserve information describing the covariance between different wavelength channels. This is another potential source of error if not properly taken into account. As has been mentioned previously, \cite{Ih2021_CorrelatedNoise} have looked at this problem but it would be of interest to combine such an analysis of the covariance with an asymmetric likelihood sampler for a more complete consideration of the errors.

Additionally, here we devote our attention to the spectrum reported for NIRSpec G395H. Other datasets exist, including those within \cite{Carter2024_wasp39b}, which report asymmetries for transmission spectra observed with other instruments. However, we choose to use the G395H dataset since it displayed the largest asymmetry of those reported. Suspecting that this may have been due to the higher resolution of the data, we also conducted the tests binned to the HST WFC3 grid and found no significant differences. Therefore, we would not expect any significant differences on other JWST spectroscopic observing modes where similar levels of asymmetry are seen. 

\section*{Acknowledgements}

This research received funding from the Science and Technology Facilities Council (STFC; grant n$^\circ$ ST/W50788X/1) and is part of the project "Interpreting exoplanet atmospheres with JWST" with file number 2024.034 of the research programme "Rekentijd nationale computersystemen" that is (partly) funded by the Netherlands Organisation for Scientific Research (NWO) under grant \url{https://doi.org/10.61686/QXVQT85756}.

Jack Davey would like to thank Dr Lorne Whiteway for his support in discussing the statistical distributions used in this project.

In the processing of these data, the following python packages were used: \textsc{NumPy} \citep{Numpy}, \textsc{SciPy} \citep{scipy}, \textsc{matplotlib} \citep{matplotlib}, \textsc{corner} \citep{corner}, \textsc{emcee} \citep{emcee} and \textsc{TauREx3} \citep{TauREx1,TauREx2,TauREx3}.

%%%%%%%%%%%%%%%%%%%%%%%%%%%%%%%%%%%%%%%%%%%%%%%%%%
\section*{Data Availability}

Simulated data used in this work can be made available upon request to the authors. The reduced data for the observations are available through a zenodo repository linked with the paper \cite{Carter2024_wasp39b}.

%%%%%%%%%%%%%%%%%%%% REFERENCES %%%%%%%%%%%%%%%%%%

% The best way to enter references is to use BibTeX:

\bibliographystyle{mnras}
\bibliography{references} % if your bibtex file is called example.bib

\begin{thebibliography}{}
\makeatletter
\relax
\def\mn@urlcharsother{\let\do\@makeother \do\$\do\&\do\#\do\^\do\_\do\%\do\~}
\def\mn@doi{\begingroup\mn@urlcharsother \@ifnextchar [ {\mn@doi@} {\mn@doi@[]}}
\def\mn@doi@[#1]#2{\def\@tempa{#1}\ifx\@tempa\@empty \href {http://dx.doi.org/#2} {doi:#2}\else \href {http://dx.doi.org/#2} {#1}\fi \endgroup}
\def\mn@eprint#1#2{\mn@eprint@#1:#2::\@nil}
\def\mn@eprint@arXiv#1{\href {http://arxiv.org/abs/#1} {{\tt arXiv:#1}}}
\def\mn@eprint@dblp#1{\href {http://dblp.uni-trier.de/rec/bibtex/#1.xml} {dblp:#1}}
\def\mn@eprint@#1:#2:#3:#4\@nil{\def\@tempa {#1}\def\@tempb {#2}\def\@tempc {#3}\ifx \@tempc \@empty \let \@tempc \@tempb \let \@tempb \@tempa \fi \ifx \@tempb \@empty \def\@tempb {arXiv}\fi \@ifundefined {mn@eprint@\@tempb}{\@tempb:\@tempc}{\expandafter \expandafter \csname mn@eprint@\@tempb\endcsname \expandafter{\@tempc}}}

\bibitem[\protect\citeauthoryear{Abel, Frommhold, Li  \& Hunt}{Abel et~al.}{2011}]{H2H2_CIA_HITRAN_Abel2011}
Abel M.,  Frommhold L.,  Li X.,   Hunt K. L.~C.,  2011, \mn@doi [J Phys Chem A] {10.1021/jp109441f}, p.~6805

\bibitem[\protect\citeauthoryear{Abel, Frommhold, Li  \& Hunt}{Abel et~al.}{2012}]{H2He_CIA_HITRAN_Abel2012}
Abel M.,  Frommhold L.,  Li X.,   Hunt K. L.~C.,  2012, \mn@doi [J Chem Phys] {10.1063/1.3676405}, 136, 044319

\bibitem[\protect\citeauthoryear{Abubekerov \& Gostev}{Abubekerov \& Gostev}{2013}]{Abubekerov2013_LimbDarkeningLaws}
Abubekerov M.~K.,  Gostev N.~Y.,  2013, \mn@doi [\mnras] {10.1093/mnras/stt575}, 432, 2216

\bibitem[\protect\citeauthoryear{Ahrer et~al.,}{Ahrer et~al.}{2023}]{NIRCam_WASP39b}
Ahrer E.-M.,  et~al., 2023, \mn@doi [Nature] {10.1038/s41586-022-05590-4}, 614, 653–658

\bibitem[\protect\citeauthoryear{Al-Refaie, Changeat, Waldmann  \& Tinetti}{Al-Refaie et~al.}{2021}]{TauREx3}
Al-Refaie A.~F.,  Changeat Q.,  Waldmann I.~P.,   Tinetti G.,  2021, \mn@doi [\apj] {10.3847/1538-4357/ac0252}, 917, 37

\bibitem[\protect\citeauthoryear{Al-Refaie, Changeat, Venot, Waldmann  \& Tinetti}{Al-Refaie et~al.}{2022}]{Al-Refaie2022_CompareChemModels}
Al-Refaie A.~F.,  Changeat Q.,  Venot O.,  Waldmann I.~P.,   Tinetti G.,  2022, \mn@doi [ApJ] {10.3847/1538-4357/ac6dcd}, 932, 123

\bibitem[\protect\citeauthoryear{Alderson, Grant  \& Wakeford}{Alderson et~al.}{2022}]{Alderson2022_ExoTiCJEDI}
Alderson L.,  Grant D.,   Wakeford H.,  2022, Exo-TiC/ExoTiC-JEDI: v0.1-beta-release, \mn@doi{10.5281/zenodo.7185855}, \url {https://doi.org/10.5281/zenodo.7185855}

\bibitem[\protect\citeauthoryear{Alderson et~al.,}{Alderson et~al.}{2023}]{NIRSpecG395H_Wasp39b}
Alderson L.,  et~al., 2023, \mn@doi [Nature] {10.1038/s41586-022-05591-3}, 614, 664–669

\bibitem[\protect\citeauthoryear{Allard, Spiegelman, Leininger  \& Molliere}{Allard et~al.}{2019}]{Na_K}
Allard N.~F.,  Spiegelman F.,  Leininger T.,   Molliere P.,  2019, \mn@doi [\aap] {10.1051/0004-6361/201935593}, 628, A120

\bibitem[\protect\citeauthoryear{{Ardévol Martínez, F.}, {Min, M.}, {Huppenkothen, D.}, {Kamp, I.}  \& {Palmer, P. I.}}{{Ardévol Martínez, F.} et~al.}{2024}]{ArdevolMartinez2024_FlopITy}
{Ardévol Martínez, F.} {Min, M.} {Huppenkothen, D.} {Kamp, I.}  {Palmer, P. I.} 2024, \mn@doi [A&A] {10.1051/0004-6361/202348367}, 681, L14

\bibitem[\protect\citeauthoryear{Arfaux \& Lavvas}{Arfaux \& Lavvas}{2024}]{Arfaux2023_CloudsAndHaze_WASP39b}
Arfaux A.,  Lavvas P.,  2024, \mn@doi [\mnras] {10.1093/mnras/stae826}, 530, 482

\bibitem[\protect\citeauthoryear{Barman}{Barman}{2007}]{Barman2007_H2O}
Barman T.,  2007, \mn@doi [ApJ] {10.1086/518736}, 661, L191

\bibitem[\protect\citeauthoryear{Barstow}{Barstow}{2020}]{Barstow2020_Clouds}
Barstow J.~K.,  2020, \mn@doi [\mnras] {10.1093/mnras/staa2219}, 497, 4183

\bibitem[\protect\citeauthoryear{{Barstow}, {Aigrain}, {Irwin}  \& {Sing}}{{Barstow} et~al.}{2017}]{Barstow2017_TenHotJupRetrieval}
{Barstow} J.~K.,  {Aigrain} S.,  {Irwin} P.~G.~J.,   {Sing} D.~K.,  2017, \mn@doi [ApJ] {10.3847/1538-4357/834/1/50}, \href {https://ui.adsabs.harvard.edu/abs/2017ApJ...834...50B} {834, 50}

\bibitem[\protect\citeauthoryear{Caldas, Leconte, Selsis, Waldmann, Bordé, Rocchetto  \& Charnay}{Caldas et~al.}{2019}]{Caldas2019_DayNightInhomog}
Caldas A.,  Leconte J.,  Selsis F.,  Waldmann I.~P.,  Bordé P.,  Rocchetto M.,   Charnay B.,  2019, \mn@doi [A&A] {10.1051/0004-6361/201834384}, 623, A161

\bibitem[\protect\citeauthoryear{{Carter} et~al.,}{{Carter} et~al.}{2024}]{Carter2024_wasp39b}
{Carter} A.~L.,  et~al., 2024, \mn@doi [Nature Astronomy] {10.1038/s41550-024-02292-x}, \href {https://ui.adsabs.harvard.edu/abs/2024NatAs...8.1008C} {8, 1008}

\bibitem[\protect\citeauthoryear{Changeat, Edwards, Waldmann  \& Tinetti}{Changeat et~al.}{2019}]{Changeat2019_TwoLayerChem}
Changeat Q.,  Edwards B.,  Waldmann I.~P.,   Tinetti G.,  2019, \mn@doi [ApJ] {10.3847/1538-4357/ab4a14}, 886, 39

\bibitem[\protect\citeauthoryear{Changeat et~al.,}{Changeat et~al.}{2022}]{changeat2022_PopStudy}
Changeat Q.,  et~al., 2022, \mn@doi [\apjs] {10.3847/1538-4365/ac5cc2}, 260, 3

\bibitem[\protect\citeauthoryear{Changeat, Ito, Al-Refaie, Yip  \& Lueftinger}{Changeat et~al.}{2024}]{Changeat2024_LightcurveRetrieval}
Changeat Q.,  Ito Y.,  Al-Refaie A.~F.,  Yip K.~H.,   Lueftinger T.,  2024, \mn@doi [AJ] {10.3847/1538-3881/ad3032}, 167, 195

\bibitem[\protect\citeauthoryear{Chubb et~al.,}{Chubb et~al.}{2021}]{ExoMol3}
Chubb K.~L.,  et~al., 2021, \mn@doi [\aap] {10.1051/0004-6361/202038350}, 646, A21

\bibitem[\protect\citeauthoryear{Constantinou \& Madhusudhan}{Constantinou \& Madhusudhan}{2022}]{Constantinou_JWSTRetrievals_MiniNeptunes}
Constantinou S.,  Madhusudhan N.,  2022, \mn@doi [\mnras] {10.1093/mnras/stac1277}, 514, 2073

\bibitem[\protect\citeauthoryear{Davey, Yip, Al-Refaie  \& Waldmann}{Davey et~al.}{2024}]{Davey2024_SpecBinning}
Davey J.~J.,  Yip K.~H.,  Al-Refaie A.~F.,   Waldmann I.~P.,  2024, \mn@doi [\mnras] {10.1093/mnras/stae2731}, 536, 2618

\bibitem[\protect\citeauthoryear{{Dressel}, {et al.,}  \& {}}{{Dressel} et~al.}{2023}]{Dressel2023_HSTWFC3Hndbk}
{Dressel} L.,  {et al.,}  {} 2023, in , WFC3 Instrument Handbook, Version 15.0.
(Baltimore: STScI)

\bibitem[\protect\citeauthoryear{Edwards et~al.,}{Edwards et~al.}{2023}]{Edwards2023_HSTPopStudy}
Edwards B.,  et~al., 2023, \mn@doi [\apjs] {10.3847/1538-4365/ac9f1a}, 269, 31

\bibitem[\protect\citeauthoryear{{Faedi} et~al.,}{{Faedi} et~al.}{2011}]{Wasp39b_Discovery}
{Faedi} F.,  et~al., 2011, \mn@doi [\aap] {10.1051/0004-6361/201116671}, \href {https://ui.adsabs.harvard.edu/abs/2011A&A...531A..40F} {531, A40}

\bibitem[\protect\citeauthoryear{Feinstein et~al.,}{Feinstein et~al.}{2023}]{NIRISS_WASP39b}
Feinstein A.~D.,  et~al., 2023, \mn@doi [Nature] {10.1038/s41586-022-05674-1}, 614, 670–675

\bibitem[\protect\citeauthoryear{Fisher \& Heng}{Fisher \& Heng}{2018}]{Fisher2018_HSTPopStudy}
Fisher C.,  Heng K.,  2018, \mn@doi [\mnras] {10.1093/mnras/sty2550}, 481, 4698

\bibitem[\protect\citeauthoryear{Foreman-Mackey}{Foreman-Mackey}{2016}]{corner}
Foreman-Mackey D.,  2016, \mn@doi [Journal of Open Source Software] {10.21105/joss.00024}, 1, 24

\bibitem[\protect\citeauthoryear{{Foreman-Mackey}, {Hogg}, {Lang}  \& {Goodman}}{{Foreman-Mackey} et~al.}{2013}]{emcee}
{Foreman-Mackey} D.,  {Hogg} D.~W.,  {Lang} D.,   {Goodman} J.,  2013, \mn@doi [\pasp] {10.1086/670067}, \href {https://ui.adsabs.harvard.edu/abs/2013PASP..125..306F} {125, 306}

\bibitem[\protect\citeauthoryear{Gao, Wakeford, Moran  \& Parmentier}{Gao et~al.}{2021}]{Gao2021_CloudsReview}
Gao P.,  Wakeford H.~R.,  Moran S.~E.,   Parmentier V.,  2021, \mn@doi [Journal of Geophysical Research: Planets] {https://doi.org/10.1029/2020JE006655}, 126, e2020JE006655

\bibitem[\protect\citeauthoryear{{Gebhard}, {Wildberger}, {Dax}, {Kofler}, {Angerhausen}, {Quanz}  \& {Sch{\"o}lkopf}}{{Gebhard} et~al.}{2025}]{Gebhard2022_FlowMatchRet}
{Gebhard} T.~D.,  {Wildberger} J.,  {Dax} M.,  {Kofler} A.,  {Angerhausen} D.,  {Quanz} S.~P.,   {Sch{\"o}lkopf} B.,  2025, \mn@doi [\aap] {10.1051/0004-6361/202451861}, \href {https://ui.adsabs.harvard.edu/abs/2025A&A...693A..42G} {693, A42}

\bibitem[\protect\citeauthoryear{{Giménez, A.}}{{Giménez, A.}}{2006}]{Gimenez2006_LightqcurveEquations}
{Giménez, A.} 2006, \mn@doi [A&A] {10.1051/0004-6361:20054445}, 450, 1231

\bibitem[\protect\citeauthoryear{Gordon et~al.,}{Gordon et~al.}{2022}]{Gordon2022_HITRAN}
Gordon I.,  et~al., 2022, \mn@doi [Journal of Quantitative Spectroscopy and Radiative Transfer] {https://doi.org/10.1016/j.jqsrt.2021.107949}, 277, 107949

\bibitem[\protect\citeauthoryear{Greene, Line, Montero, Fortney, Lustig-Yaeger  \& Luther}{Greene et~al.}{2016}]{Greene2016_JWSTTransmissionSpecOverview}
Greene T.~P.,  Line M.~R.,  Montero C.,  Fortney J.~J.,  Lustig-Yaeger J.,   Luther K.,  2016, \mn@doi [ApJ] {10.3847/0004-637X/817/1/17}, 817, 17

\bibitem[\protect\citeauthoryear{Guzmán-Mesa et~al.,}{Guzmán-Mesa et~al.}{2020}]{Guzmán-Mesa_JWST_InformationContent_WarmNep}
Guzmán-Mesa A.,  et~al., 2020, \mn@doi [AJ] {10.3847/1538-3881/ab9176}, 160, 15

\bibitem[\protect\citeauthoryear{Harris et~al.,}{Harris et~al.}{2020}]{Numpy}
Harris C.~R.,  et~al., 2020, \mn@doi [Nature] {10.1038/s41586-020-2649-2}, 585, 357

\bibitem[\protect\citeauthoryear{Hunter}{Hunter}{2007}]{matplotlib}
Hunter J.~D.,  2007, \mn@doi [Computing in Science & Engineering] {10.1109/MCSE.2007.55}, 9, 90

\bibitem[\protect\citeauthoryear{Ih \& Kempton}{Ih \& Kempton}{2021}]{Ih2021_CorrelatedNoise}
Ih J.,  Kempton E. M.-R.,  2021, \mn@doi [AJ] {10.3847/1538-3881/ac173b}, 162, 237

\bibitem[\protect\citeauthoryear{Kitzmann, Heng, Oreshenko, Grimm, Apai, Bowler, Burgasser  \& Marley}{Kitzmann et~al.}{2020}]{Kitzmann2020_ErrInflationHELIOS}
Kitzmann D.,  Heng K.,  Oreshenko M.,  Grimm S.~L.,  Apai D.,  Bowler B.~P.,  Burgasser A.~J.,   Marley M.~S.,  2020, \mn@doi [ApJ] {10.3847/1538-4357/ab6d71}, 890, 174

\bibitem[\protect\citeauthoryear{Kreidberg}{Kreidberg}{2015}]{Kreidberg2015_Batman}
Kreidberg L.,  2015, \mn@doi [Publications of the Astronomical Society of the Pacific] {10.1086/683602}, 127, 1161

\bibitem[\protect\citeauthoryear{Kreidberg et~al.,}{Kreidberg et~al.}{2014}]{Kreidberg2014_CloudsSuperEarth}
Kreidberg L.,  et~al., 2014, \mn@doi [Nature] {10.1038/nature12888}, 505, 69–72

\bibitem[\protect\citeauthoryear{Lacy \& Burrows}{Lacy \& Burrows}{2020}]{Lacy2020_DayNightInhomog}
Lacy B.~I.,  Burrows A.,  2020, \mn@doi [ApJ] {10.3847/1538-4357/abc01c}, 905, 131

\bibitem[\protect\citeauthoryear{Li, Gordon, Rothman, Tan, Hu, Kassi, Campargue  \& Medvedev}{Li et~al.}{2015}]{CO}
Li G.,  Gordon I.~E.,  Rothman L.~S.,  Tan Y.,  Hu S.-M.,  Kassi S.,  Campargue A.,   Medvedev E.~S.,  2015, \mn@doi [\apjs] {10.1088/0067-0049/216/1/15}, 216, 15

\bibitem[\protect\citeauthoryear{Line et~al.,}{Line et~al.}{2013}]{Line2013_Prior}
Line M.~R.,  et~al., 2013, \mn@doi [ApJ] {10.1088/0004-637X/775/2/137}, 775, 137

\bibitem[\protect\citeauthoryear{Line, Teske, Burningham, Fortney  \& Marley}{Line et~al.}{2015}]{Line2015_ErrInflation}
Line M.~R.,  Teske J.,  Burningham B.,  Fortney J.~J.,   Marley M.~S.,  2015, \mn@doi [ApJ] {10.1088/0004-637X/807/2/183}, 807, 183

\bibitem[\protect\citeauthoryear{Lueber, Novais, Fisher  \& Heng}{Lueber et~al.}{2024}]{Lueber2024_InfoContentJWSTWASP39b}
Lueber A.,  Novais A.,  Fisher C.,   Heng K.,  2024, \mn@doi [A&A] {10.1051/0004-6361/202348802}, 687, A110

\bibitem[\protect\citeauthoryear{{Lueber}, {Karchev}, {Fisher}, {Heim}, {Trotta}  \& {Heng}}{{Lueber} et~al.}{2025}]{Lueber2025_FASTER_LFIRetrievals}
{Lueber} A.,  {Karchev} K.,  {Fisher} C.,  {Heim} M.,  {Trotta} R.,   {Heng} K.,  2025, \mn@doi [arXiv e-prints] {10.48550/arXiv.2502.18045}, \href {https://ui.adsabs.harvard.edu/abs/2025arXiv250218045L} {p. arXiv:2502.18045}

\bibitem[\protect\citeauthoryear{Ma, Ito, Al-Refaie, Changeat, Edwards  \& Tinetti}{Ma et~al.}{2023}]{Ma2023_YunMaClouds}
Ma S.,  Ito Y.,  Al-Refaie A.~F.,  Changeat Q.,  Edwards B.,   Tinetti G.,  2023, \mn@doi [ApJ] {10.3847/1538-4357/acf8ca}, 957, 104

\bibitem[\protect\citeauthoryear{MacDonald \& Madhusudhan}{MacDonald \& Madhusudhan}{2017}]{MacDonald2017_PatchyClouds_HD209}
MacDonald R.~J.,  Madhusudhan N.,  2017, \mn@doi [\mnras] {10.1093/mnras/stx804}, 469, 1979

\bibitem[\protect\citeauthoryear{{Maciejewski} et~al.,}{{Maciejewski} et~al.}{2016}]{Maciejewski2016_WASP39b_OrbitalParams}
{Maciejewski} G.,  et~al., 2016, \mn@doi [\actaa] {10.48550/arXiv.1603.03268}, \href {https://ui.adsabs.harvard.edu/abs/2016AcA....66...55M} {66, 55}

\bibitem[\protect\citeauthoryear{Madhusudhan}{Madhusudhan}{2018}]{RetrievalSummaryChap}
Madhusudhan N.,  2018, Atmospheric Retrieval of Exoplanets.
p. 2153–2182, \mn@doi{10.1007/978-3-319-55333-7_104}

\bibitem[\protect\citeauthoryear{{Madhusudhan} \& {Seager}}{{Madhusudhan} \& {Seager}}{2009}]{FirstRetrievals}
{Madhusudhan} N.,  {Seager} S.,  2009, \mn@doi [ApJ] {10.1088/0004-637X/707/1/24}, \href {https://ui.adsabs.harvard.edu/abs/2009ApJ...707...24M} {707, 24}

\bibitem[\protect\citeauthoryear{Mandel \& Agol}{Mandel \& Agol}{2002}]{Mandel2002_LightcurveEqns}
Mandel K.,  Agol E.,  2002, \mn@doi [ApJ] {10.1086/345520}, 580, L171

\bibitem[\protect\citeauthoryear{Morvan, Tsiaras, Nikolaou  \& Waldmann}{Morvan et~al.}{2021}]{Morvan2021_PyLightCurve-Torch}
Morvan M.,  Tsiaras A.,  Nikolaou N.,   Waldmann I.~P.,  2021, \mn@doi [Publications of the Astronomical Society of the Pacific] {10.1088/1538-3873/abe6e8}, 133, 034505

\bibitem[\protect\citeauthoryear{Nixon \& Madhusudhan}{Nixon \& Madhusudhan}{2020}]{Nixon2020_ML4Retrievals}
Nixon M.~C.,  Madhusudhan N.,  2020, \mn@doi [\mnras] {10.1093/mnras/staa1150}, 496, 269

\bibitem[\protect\citeauthoryear{Pacetti et~al.,}{Pacetti et~al.}{2022}]{Pacetti2022_DiskChemDiversity}
Pacetti E.,  et~al., 2022, \mn@doi [ApJ] {10.3847/1538-4357/ac8b11}, \href {https://ui.adsabs.harvard.edu/abs/2022ApJ...937...36P} {937, 36}

\bibitem[\protect\citeauthoryear{Pinhas, Rackham, Madhusudhan  \& Apai}{Pinhas et~al.}{2018}]{Pinhas2018_StellarContamination}
Pinhas A.,  Rackham B.~V.,  Madhusudhan N.,   Apai D.,  2018, \mn@doi [\mnras] {10.1093/mnras/sty2209}, 480, 5314

\bibitem[\protect\citeauthoryear{Polyansky, Kyuberis, Zobov, Tennyson, Yurchenko  \& Lodi}{Polyansky et~al.}{2018}]{H2O}
Polyansky O.~L.,  Kyuberis A.~A.,  Zobov N.~F.,  Tennyson J.,  Yurchenko S.~N.,   Lodi L.,  2018, \mn@doi [\mnras] {10.1093/mnras/sty1877}, 480, 2597

\bibitem[\protect\citeauthoryear{Richard et~al.,}{Richard et~al.}{2012}]{RICHARD2012_HITRAN_CIA}
Richard C.,  et~al., 2012, \mn@doi [Journal of Quantitative Spectroscopy and Radiative Transfer] {https://doi.org/10.1016/j.jqsrt.2011.11.004}, 113, 1276

\bibitem[\protect\citeauthoryear{Rothman et~al.,}{Rothman et~al.}{2013}]{CIA_Hitran_2013}
Rothman L.,  et~al., 2013, \mn@doi [Journal of Quantitative Spectroscopy and Radiative Transfer] {https://doi.org/10.1016/j.jqsrt.2013.07.002}, 130, 4

\bibitem[\protect\citeauthoryear{Rustamkulov et~al.,}{Rustamkulov et~al.}{2023}]{NIRSpecPRISM_Wasp39b}
Rustamkulov Z.,  et~al., 2023, \mn@doi [Nature] {10.1038/s41586-022-05677-y}, 614, 659–663

\bibitem[\protect\citeauthoryear{{Sanders} \& {Evans}}{{Sanders} \& {Evans}}{2020}]{Sanders2020_AsymDists}
{Sanders} J.~L.,  {Evans} N.~W.,  2020, \mn@doi [\mnras] {10.1093/mnras/staa2860}, \href {https://ui.adsabs.harvard.edu/abs/2020MNRAS.499.5806S} {499, 5806}

\bibitem[\protect\citeauthoryear{Schleich, Boro~Saikia, Changeat, Güdel, Voigt  \& Waldmann}{Schleich et~al.}{2024}]{Schleich2024_TpProfileComplexity}
Schleich S.,  Boro~Saikia S.,  Changeat Q.,  Güdel M.,  Voigt A.,   Waldmann I.,  2024, \mn@doi [A&A] {10.1051/0004-6361/202451845}, 690, A336

\bibitem[\protect\citeauthoryear{Sing et~al.,}{Sing et~al.}{2016}]{Sing2016_TenHotJupiters}
Sing D.~K.,  et~al., 2016, \mn@doi [Nature] {10.1038/nature16068}, 529, 59–62

\bibitem[\protect\citeauthoryear{{Stassun} et~al.,}{{Stassun} et~al.}{2019}]{Stassun2019_WASP39_StellarParams}
{Stassun} K.~G.,  et~al., 2019, \mn@doi [AJ] {10.3847/1538-3881/ab3467}, \href {https://ui.adsabs.harvard.edu/abs/2019AJ....158..138S} {158, 138}

\bibitem[\protect\citeauthoryear{Swain, Vasisht  \& Tinetti}{Swain et~al.}{2008}]{Swain2008_CH4}
Swain M.~R.,  Vasisht G.,   Tinetti G.,  2008, \mn@doi [Nature] {10.1038/nature06823}, 452, 329–331

\bibitem[\protect\citeauthoryear{Taylor, Parmentier, Irwin, Aigrain, Lee  \& Krissansen-Totton}{Taylor et~al.}{2020}]{Taylor2020_BiasEmissionSpecModel}
Taylor J.,  Parmentier V.,  Irwin P. G.~J.,  Aigrain S.,  Lee E.,   Krissansen-Totton J.,  2020, \mn@doi [\mnras] {10.1093/mnras/staa552}, 493, 4342

\bibitem[\protect\citeauthoryear{Taylor et~al.,}{Taylor et~al.}{2023}]{Taylor2023_NIRISS_WASP96b}
Taylor J.,  et~al., 2023, \mn@doi [\mnras] {10.1093/mnras/stad1547}, 524, 817

\bibitem[\protect\citeauthoryear{{Tennyson} \& {Yurchenko}}{{Tennyson} \& {Yurchenko}}{2012}]{ExoMol1}
{Tennyson} J.,  {Yurchenko} S.~N.,  2012, \mn@doi [\mnras] {10.1111/j.1365-2966.2012.21440.x}, \href {https://ui.adsabs.harvard.edu/abs/2012MNRAS.425...21T} {425, 21}

\bibitem[\protect\citeauthoryear{{Tennyson} et~al.,}{{Tennyson} et~al.}{2016}]{ExoMol2}
{Tennyson} J.,  et~al., 2016, \mn@doi [Journal of Molecular Spectroscopy] {10.1016/j.jms.2016.05.002}, \href {https://ui.adsabs.harvard.edu/abs/2016JMoSp.327...73T} {327, 73}

\bibitem[\protect\citeauthoryear{Thompson et~al.,}{Thompson et~al.}{2024}]{Thompson2024_StellarContamination}
Thompson A.,  et~al., 2024, \mn@doi [ApJ] {10.3847/1538-4357/ad0369}, 960, 107

\bibitem[\protect\citeauthoryear{Tinetti et~al.,}{Tinetti et~al.}{2007}]{Tinetti2007_H2O}
Tinetti G.,  et~al., 2007, \mn@doi [Nature] {10.1038/nature06002}, 448, 169–171

\bibitem[\protect\citeauthoryear{Tinetti et~al.,}{Tinetti et~al.}{2018}]{Tinetti2018_ArielOverview}
Tinetti G.,  et~al., 2018, \mn@doi [Experimental Astronomy] {10.1007/s10686-018-9598-x}, 46, 135–209

\bibitem[\protect\citeauthoryear{Tsiaras et~al.,}{Tsiaras et~al.}{2018}]{Tsiaras2018_HSTPopStudy}
Tsiaras A.,  et~al., 2018, \mn@doi [AJ] {10.3847/1538-3881/aaaf75}, 155, 156

\bibitem[\protect\citeauthoryear{Underwood, Tennyson, Yurchenko, Huang, Schwenke, Lee, Clausen  \& Fateev}{Underwood et~al.}{2016}]{SO2}
Underwood D.~S.,  Tennyson J.,  Yurchenko S.~N.,  Huang X.,  Schwenke D.~W.,  Lee T.~J.,  Clausen S.,   Fateev A.,  2016, \mn@doi [\mnras] {10.1093/mnras/stw849}, 459, 3890

\bibitem[\protect\citeauthoryear{Vasist, Rozet, Absil, Mollière, Nasedkin  \& Louppe}{Vasist et~al.}{2023}]{Vasist2023_NPERetrieval}
Vasist M.,  Rozet F.,  Absil O.,  Mollière P.,  Nasedkin E.,   Louppe G.,  2023, \mn@doi [A&A] {10.1051/0004-6361/202245263}, 672, A147

\bibitem[\protect\citeauthoryear{Virtanen et~al.,}{Virtanen et~al.}{2020}]{scipy}
Virtanen P.,  et~al., 2020, \mn@doi [Nature Methods] {10.1038/s41592-019-0686-2}, \href {https://rdcu.be/b08Wh} {17, 261}

\bibitem[\protect\citeauthoryear{{Wakeford} et~al.,}{{Wakeford} et~al.}{2018}]{HST_WFC3_WASP39b}
{Wakeford} H.~R.,  et~al., 2018, \mn@doi [AJ] {10.3847/1538-3881/aa9e4e}, \href {https://ui.adsabs.harvard.edu/abs/2018AJ....155...29W} {155, 29}

\bibitem[\protect\citeauthoryear{{Waldmann}, {Tinetti}, {Rocchetto}, {Barton}, {Yurchenko}  \& {Tennyson}}{{Waldmann} et~al.}{2015a}]{TauREx1}
{Waldmann} I.~P.,  {Tinetti} G.,  {Rocchetto} M.,  {Barton} E.~J.,  {Yurchenko} S.~N.,   {Tennyson} J.,  2015a, \mn@doi [ApJ] {10.1088/0004-637X/802/2/107}, \href {https://ui.adsabs.harvard.edu/abs/2015ApJ...802..107W} {802, 107}

\bibitem[\protect\citeauthoryear{{Waldmann}, {Rocchetto}, {Tinetti}, {Barton}, {Yurchenko}  \& {Tennyson}}{{Waldmann} et~al.}{2015b}]{TauREx2}
{Waldmann} I.~P.,  {Rocchetto} M.,  {Tinetti} G.,  {Barton} E.~J.,  {Yurchenko} S.~N.,   {Tennyson} J.,  2015b, \mn@doi [ApJ] {10.1088/0004-637X/813/1/13}, \href {https://ui.adsabs.harvard.edu/abs/2015ApJ...813...13W} {813, 13}

\bibitem[\protect\citeauthoryear{Yip, Tsiaras, Waldmann  \& Tinetti}{Yip et~al.}{2020}]{Yip2020_LightcurveRetrieval}
Yip K.~H.,  Tsiaras A.,  Waldmann I.~P.,   Tinetti G.,  2020, \mn@doi [AJ] {10.3847/1538-3881/abaabc}, 160, 171

\bibitem[\protect\citeauthoryear{Yip, Changeat, Nikolaou, Morvan, Edwards, Waldmann  \& Tinetti}{Yip et~al.}{2021}]{Yip2021_DeepLearningModels}
Yip K.~H.,  Changeat Q.,  Nikolaou N.,  Morvan M.,  Edwards B.,  Waldmann I.~P.,   Tinetti G.,  2021, \mn@doi [AJ] {10.3847/1538-3881/ac1744}, 162, 195

\bibitem[\protect\citeauthoryear{Yurchenko, Mellor, Freedman  \& Tennyson}{Yurchenko et~al.}{2020}]{CO2}
Yurchenko S.~N.,  Mellor T.~M.,  Freedman R.~S.,   Tennyson J.,  2020, \mn@doi [\mnras] {10.1093/mnras/staa1874}, 496, 5282

\bibitem[\protect\citeauthoryear{{Zhu} \& {Dong}}{{Zhu} \& {Dong}}{2021}]{Zhu2021_ExoStatisticsReview}
{Zhu} W.,  {Dong} S.,  2021, \mn@doi [\araa] {10.1146/annurev-astro-112420-020055}, \href {https://ui.adsabs.harvard.edu/abs/2021ARA&A..59..291Z} {59, 291}

\bibitem[\protect\citeauthoryear{Öberg, Murray-Clay  \& Bergin}{Öberg et~al.}{2011}]{Oberg2011_IceLines}
Öberg K.~I.,  Murray-Clay R.,   Bergin E.~A.,  2011, \mn@doi [The Astrophysical Journal Letters] {10.1088/2041-8205/743/1/L16}, 743, L16

\makeatother
\end{thebibliography}

% Alternatively you could enter them by hand, like this:
% This method is tedious and prone to error if you have lots of references
%\begin{thebibliography}{99}
%\bibitem[\protect\citeauthoryear{Author}{2012}]{Author2012}
%Author A.~N., 2013, Journal of Improbable Astronomy, 1, 1
%\bibitem[\protect\citeauthoryear{Others}{2013}]{Others2013}
%Others S., 2012, Journal of Interesting Stuff, 17, 198
%\end{thebibliography}

%%%%%%%%%%%%%%%%%%%%%%%%%%%%%%%%%%%%%%%%%%%%%%%%%%

%%%%%%%%%%%%%%%%% APPENDICES %%%%%%%%%%%%%%%%%%%%%

\appendix

\section{Fitting An Exponentially Modified Normal}
\label{app:FitAsymDist}

\begin{figure}
    \includegraphics[width=\columnwidth]{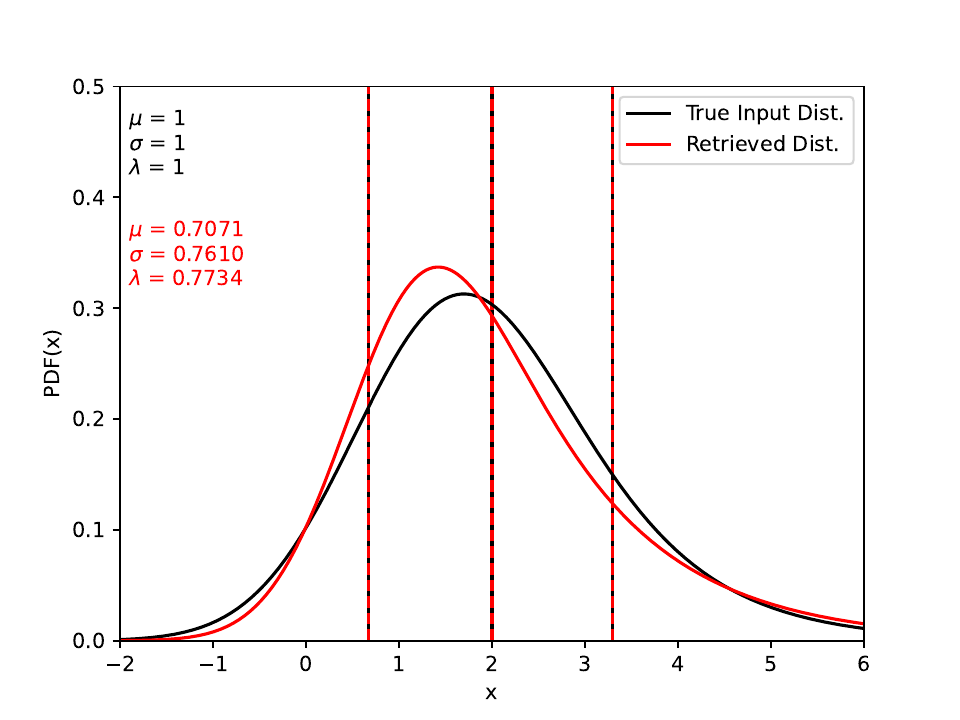}
    \caption{Degeneracy in the exponentially modified normal distribution: the black $(\mu, \sigma, \lambda) = (1, 1, 1)$ and red $(\mu, \sigma, \lambda) = (0.7071, 0.7610, 0.7734)$ have the same mean and 16\textsuperscript{th} and 84\textsuperscript{th} percentiles but they are different distributions.}
    \label{fig:DegenSolns}
\end{figure}

Given that our retrieval results highlight the importance of knowing the specific shape of the noise distribution we attempt to fit an asymmetric distribution based on data as it is currently reported in the literature. Most often, authors will offer a central value (either the mean or median of their posterior distribution) and then either a single error or a pair of error bars. In the symmetric case, two values, the median (equivalent to the mean) and the variance, are sufficient to fully describe the Gaussian distribution under investigation. If only one value for error is given, it is common to assume this to be the true scenario. In the asymmetric case, adding a third value to our reported results creates ambiguity as there are many viable asymmetric distributions which share the same set of summary statistics. There is no commonly assumed choice in this instance.

Attempts were made to reproduce the observed levels of asymmetry using several different asymmetric distributions (e.g. a skewed normal, an exponentially modified normal, Johnson's SU) and under both the assumption of the central value reported as the mean and as the median. In both cases, these well-defined distributions were found to be insufficient to reproduce the observed levels of asymmetry. Furthermore, with the assumption of the central value being the mean and the distribution being the exponentially modified normal, we found issues with degenerate solutions. 

For the purposes of this investigation, we assume that we know the shape of the asymmetric distribution to be that of an exponentially modified Gaussian distribution. Assuming our reported data provides the mean and the 84\textsuperscript{th} and 16\textsuperscript{th} percentiles, we attempt to fit the full PDF (defined in equation~\ref{eq:ExpoNormalPDF}).

The exponentially modified normal distribution is constructed as a convolution of a normal distribution and an exponential curve, damping the positive side of the distribution. It can be characterised by three values, the mean and variance of the underlying Gaussian component ($\mathrm{\mu}$ and $\mathrm{\sigma}$ respectively) and $\mathrm{\lambda}$, a parameter that modifies the damping of the exponential term. The corresponding cumulative density function (CDF) for this distribution is
\begin{multline}
    \mathrm{CDF(x;\mu,\sigma^{2},\lambda) = \Phi(x,\mu,\sigma)-}\\\mathrm{\frac{1}{2} \exp\left({\frac{\lambda}{2}(2\mu+\lambda\sigma^{2}-2x)}\right)\left(1+erf\left(\frac{x-\mu-\lambda\sigma^{2}}{\sqrt{2}\sigma}\right) \right)}    
\end{multline}
where $\mathrm{\Phi(\frac{x-\mu}{\sigma}) = \frac{1}{2} \left(1 + erf\left(\frac{x-\mu}{\sigma\sqrt{2}}\right)\right)}$ gives the CDF for a standard Gaussian distribution. The mean of the distribution is
\begin{equation}
    \mathrm{X} = \mu + \frac{1}{\lambda}.
\end{equation}

For a spectral data point, $\mathrm{X^{\sigma_{\uparrow}}_{\sigma_{\downarrow}}}$ we attempt to fit values for the $\mu$, $\sigma$ and $\lambda$ using Newton's method and the known equations of $\mathrm{CDF(\sigma_{\uparrow};\mu,\sigma^{2},\lambda) = 0.84}$, $\mathrm{CDF(\sigma_{\downarrow};\mu,\sigma^{2},\lambda) = 0.16}$ and $\mathrm{X} = \mu + \frac{1}{\lambda}$. However, in this analysis we are heavily restricted. If we operate under the broad assumption that our choice of asymmetric distribution is perfect, even in these cases, we find degenerate solutions for $\mu,~\sigma$ and $\lambda$ giving the same values for our three condition equations. 

In figure \ref{fig:DegenSolns} we display an example of a $(\mu, \sigma, \lambda)~=~(1, 1, 1)$ distribution where the fitting converges on a solution of $(0.707, 0.761, 0.773)$ but the summary statistics are fit perfectly. Thus, we advocate for more information being provided when reporting spectra. The full set of posterior distributions would allow us to operate with complete confidence in the shape of our noise distribution at the spectrum fitting stage. We see this as a necessary step if we reach a situation in which we expect significant deviations from the Gaussian assumption.

\section{Consistency of Repeated Experiments}
\label{app:RepeatRetrievals}

\begin{figure*}
    \includegraphics[width=\textwidth]{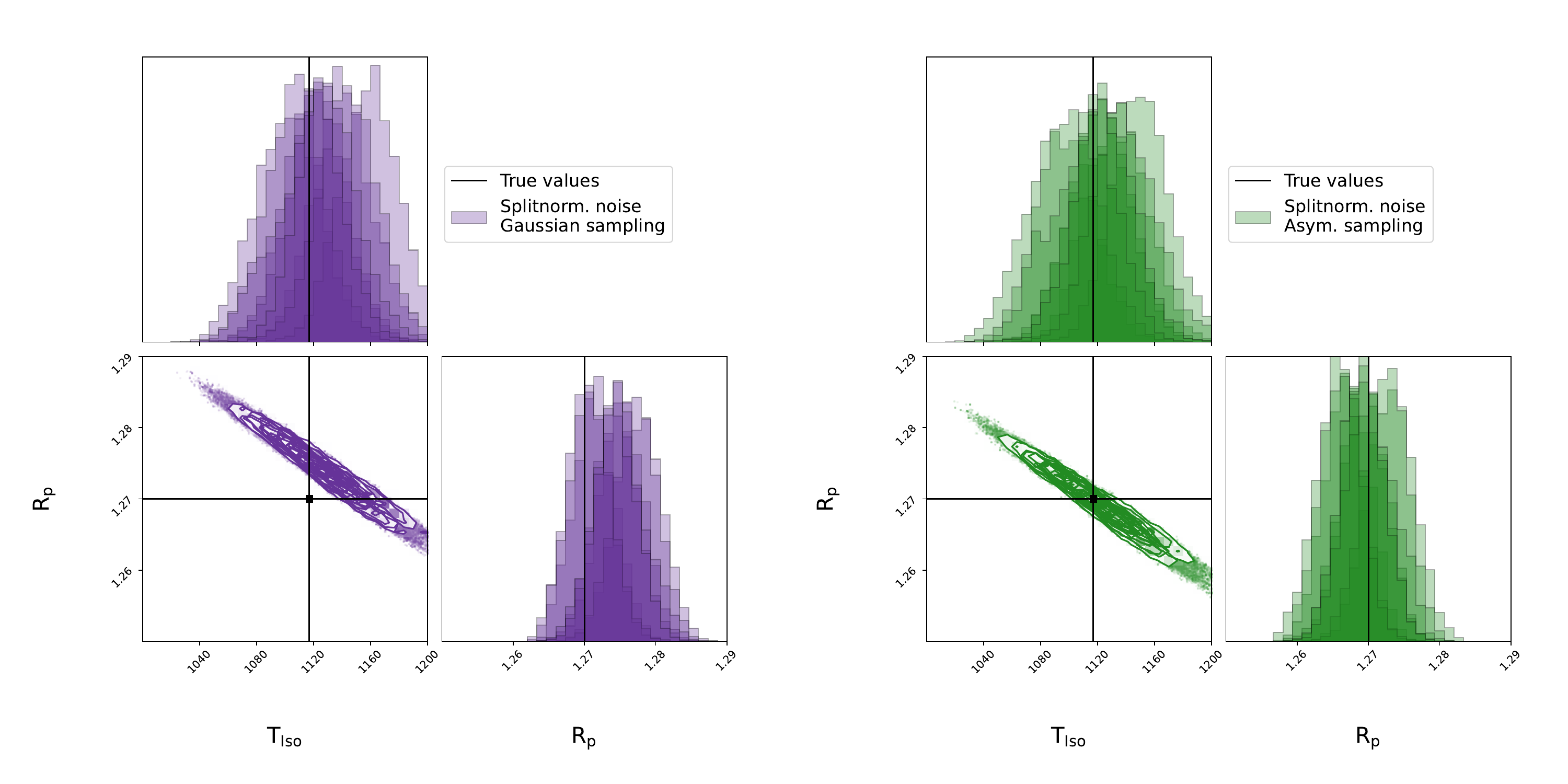}
    \caption{Retrievals for nine independent noise instances on the simulation of WASP-39\,b across the JWST NIRSpec G395H wavelength grid. The simulation had noise added according to split normal distributions with a +50\% asymmetry on every point where the width of the noise distribution was scaled by a factor of 1.5 compared to the observed errors in \protect\cite{Carter2024_wasp39b}. On the left are the results when sampling with a Gaussian likelihood while on the right are those when sampling with the asymmetric likelihood.}
    \label{fig:Repeats}
\end{figure*}

\begin{figure*}
    \includegraphics[width=\textwidth]{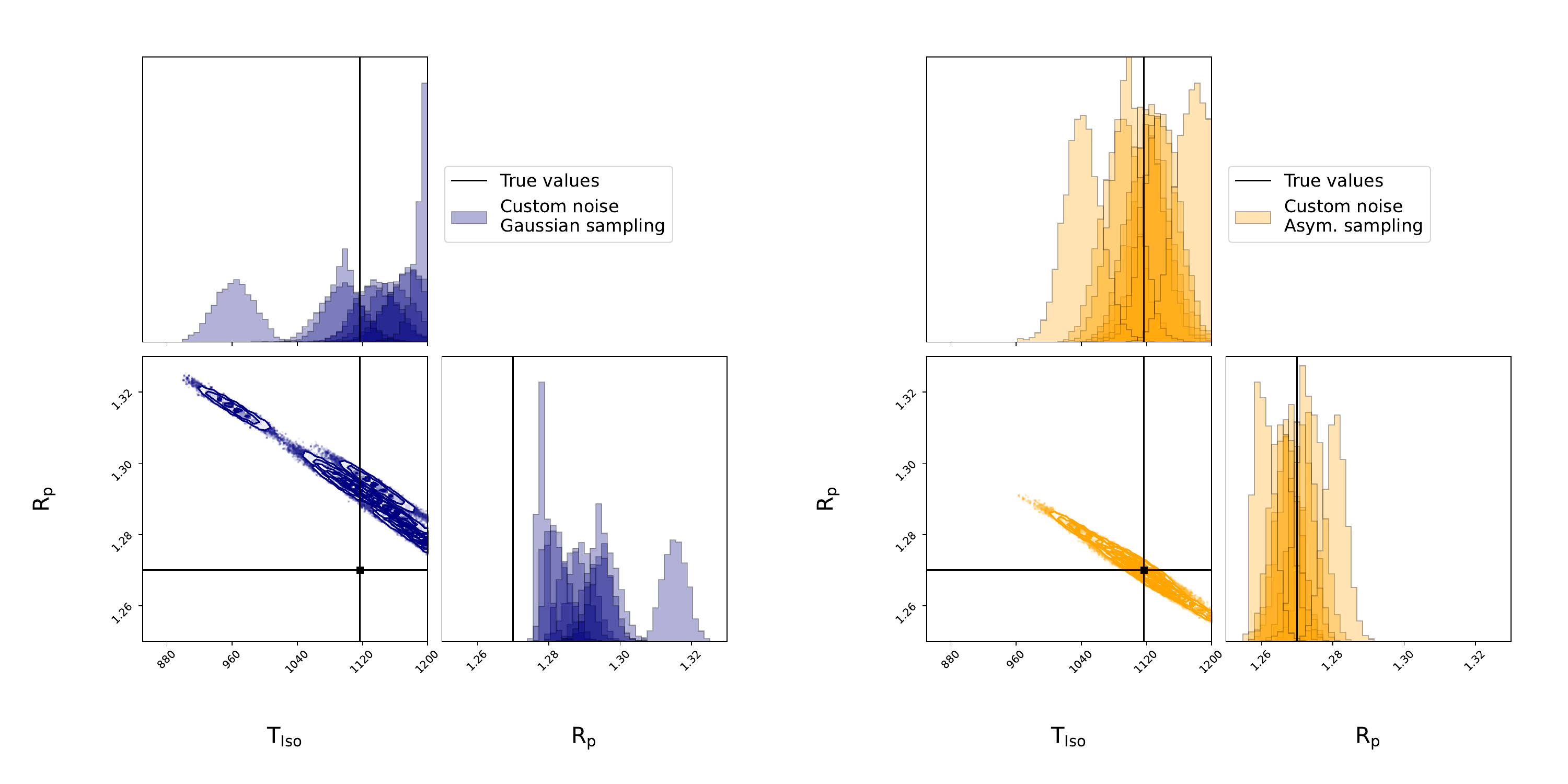}
    \caption{Retrievals for nine independent noise instances on the simulation of WASP-39\,b across the JWST NIRSpec G395H wavelength grid. The simulation had noise added according to custom distributions with a +125\% asymmetry on every point where the width of the noise distribution was scaled by a factor of 1.5 compared to the observed errors in \protect\cite{Carter2024_wasp39b}. On the left are the results when sampling with a Gaussian likelihood while on the right are those when sampling with the asymmetric likelihood.}
    \label{fig:Repeats_custom}
\end{figure*}

In this section we provide an investigation of the consistency of our retrievals. The aim here is to ensure that we are not biased by a single instance of the random noise. In figure \ref{fig:Repeats}, we use the case of a +50\% asymmetry and generate nine different datasets. Each dataset has noise added according to a split normal distribution at each point with the error bars above and below set by 1.5 times the reported errors (scaling the upper error such that +50\% asymmetry is achieved). Figure \ref{fig:Repeats_custom} shows the same test applied to the custom noise distribution with a +125\% asymmetry.

In figure \ref{fig:Repeats}, we plot the data from the nine instances of split normal noise on the same corner plots but separate by sampling method for clarity. These data show clear consistency in their general shape and position for each parameter in each sampling scheme. In figure \ref{fig:Repeats_custom} we show the results of the retrievals for instances of the custom noise. We note less consistency in the Gaussian retrievals than in the retrievals with asymmetric sampling. This further emphasises the increased reliability of predictions made by sampling which correctly characterises the noise distributions compared to those which don't. 


\section{Computing the two-sample Kolmogorov-Smirnov test}
\label{app:ks_2samp}

To provide a quantitative measure of the agreement between each pair of one-dimensional marginal distributions from the two sampling likelihoods on each noisy dataset for the sine wave model, here (in table \ref{table:KS_SineWave}) we offer the two-sample Kolmogorov-Smirnov test (KS-test). When this test is far from zero, it means that the two samples are unlikely to have been drawn from the same underlying distribution. The KS-test is implemented through its inclusion in the \textsc{SciPy} package.

\begin{table}
    \centering
    \begin{tabular}{c|c|c}
    \toprule\toprule
         \textbf{Noise case} & \textbf{a}  & \textbf{c} \\
         \midrule
         Gaussian  & 0.008 & 0.007 \\ \\
         Asymmetric Shift  &  0.188 & 0.999 \\ \\
         Asymmetric Compression  & 0.985 & 0.318 \\ \\
         Asymmetric with Red & 0.018 & 0.993 \\
         Noise Scaling  & & \\
    \end{tabular}
    \caption{Results from the two-sample KS-test comparing the one-dimensional marginal distributions for Gaussian and asymmetric likelihood retrievals on the noisy sine wave spectra.}
    \label{table:KS_SineWave}
\end{table}

%%%%%%%%%%%%%%%%%%%%%%%%%%%%%%%%%%%%%%%%%%%%%%%%%%

% Don't change these lines
\bsp	% typesetting comment
\label{lastpage}
\end{document}